\documentclass[preprint,12pt]{elsarticle}

\usepackage[numbers]{natbib}
\usepackage[english]{babel}
\usepackage{placeins}
\usepackage{amssymb,amsmath,empheq,xcolor}
\usepackage{wasysym}
\usepackage{placeins} 
\usepackage{xcolor} 
\usepackage{xspace} 

\usepackage{mathtools}
\usepackage{siunitx}
\usepackage{diffcoeff}
\usepackage{upgreek}
\usepackage[normalem]{ulem}
\usepackage[utf8]{inputenc}
\usepackage{float}
\usepackage[normalem]{ulem}

\usepackage{lineno}
\usepackage{graphicx}

\newcommand{\figref}[1]{Fig.~\ref{#1}}
\newcommand{\Figref}[1]{Figure~\ref{#1}}	
\newcommand{\secref}[1]{Section~\ref{#1}}

\newcommand{\ArCOtwo}{Ar-CO$_2$ (90-10)\xspace}

\newcommand{\TtwoK}{Ar-CF$_4$-$i$C$_4$H$_{10}$ (95-3-2)\xspace}

\journal{Nuclear Instruments and Methods in Physics Research Section A}

\begin{document}

\begin{frontmatter}



\title{The HYDRA pion-tracker for hypernuclei studies at R$^3$B}

\author[aff1,aff2]{Lian-Cheng Ji}
\author[aff1]{Uwe Bonnes}
\author[aff8]{Mikołaj Ćwiok}
\author[aff1]{Meytal Duer}
\author[aff1]{Alexandru Enciu}
\author[aff1,aff3,aff4]{Piotr Gasik}
\author[aff3]{Joerg Hehner}
\author[aff1,aff2]{Alexandre Obertelli}
\author[aff5,aff6]{Shinsuke Ota}
\author[aff3]{Valerii Panin}
\author[aff7]{Jérôme Pibernat}
\author[aff1,aff3]{Dominic Rossi}
\author[aff3,aff4]{Haik Simon}
\author[aff1]{Yelei Sun}
\author[aff1]{Simone Velardita}
\author[aff1]{Frank Wienholtz}
\author[aff8]{Marcin Zaremba}

\affiliation[aff1]{organization={Technische Universität Darmstadt, Fachbereich Physik},
            city={Darmstadt},
            country={Germany}}
\affiliation[aff2]{organization={Helmholtz Forschungsakademie Hessen für FAIR (HFHF)},
            city={Frankfurt},
            country={Germany}}
\affiliation[aff3]{organization={GSI Helmholtzzentrum für Schwerionenforschung GmbH},
            city={Darmstadt},
            country={Germany}}
\affiliation[aff4]{organization={FAIR Facility for Antiproton and Ion Research in Europe GmbH},
            city={Darmstadt},
            country={Germany}}
\affiliation[aff5]{organization={Research Center for Nuclear Physics, Osaka University},
            city={Ibaraki},
            country={Japan}}
\affiliation[aff6]{organization={Center for Nuclear Study, the University of Tokyo},
            city={Saitama},
            country={Japan}}
\affiliation[aff7]{organization={Centre d'Etudes Nucléaires de Bordeaux-Gradignan - Université Bordeaux I - CNRS/IN2P3},
            city={Gradignan},
            country={France}}
\affiliation[aff8]{organization={University of Warsaw, Faculty of Physics},
            city={Warsaw},
            country={Poland}}

\begin{abstract}
    The HYpernuclei-Decay at R$^3$B Apparatus (HYDRA) tracker is a novel time projection chamber combined with a plastic scintillator wall for timing and trigger purposes. This detector is a low radiation length tracker dedicated to measuring pions from the weak decay of light hypernuclei produced from ion-ion collisions at few GeV/nucleon in the magnetic field of the large-acceptance dipole magnet GLAD at the Reactions with Relativistic Radioactive Beams (R$^3$B) experiment at GSI-FAIR. In this paper, we describe the design of the detector and provide the results of its first characterizations. 
\end{abstract}

\begin{keyword}
Hypernuclei \sep Ion-ion collisions \sep Time projection chamber
\end{keyword}

\end{frontmatter}


\section{Introduction}
    \noindent The current knowledge concerning nuclear interactions, nuclear physics, and the nuclear equation of state relies primarily on the study of nuclei that are composed of only up (${u}$) and down (${d}$) valence quarks. Yet the presence of the strange (${s}$) quark, the third lightest quark within the Standard Model, allows for the inclusion of strangeness in nuclear matter. In particular, hyperons, the baryons comprising at least one ${s}$ quark, can form bound systems with nucleons and create short-lived hypernuclei~\cite{Danysz1953}. For example, single-$\rm \Lambda$-hypernuclei contain nucleons and the lightest hyperon, the $\rm \Lambda$ baryon (${u}$,${d}$,${s}$). Hypernuclear studies shed a new light on the world of nuclear interaction \cite{Meiner2017} and traditional nuclei by revealing new symmetries and phenomena produced due to the additional strangeness degree of freedom, and have become the most important means to explore the hyperon-nucleon interactions \cite{Petschauer2020}, though only few scattering data exist to constrain \cite{Haidenbauer2013,Haidenbauer2020,ALICECollaboration2020}. While about 3,500 bound nuclei have been discovered \cite{Erler2012}, in the case of single- and double-$\rm \Lambda$-hypernuclei, only 41 have been synthesized so far \cite{Mainz2025}, often with little data and uncontrolled systematic uncertainties, leading sometimes to debated results, despite decades of experimental efforts. \\

    \noindent The scarcity of experimental data and large uncertainties stem from two inherent challenges. Firstly, the production of hypernuclei is constrained by the necessity of introducing a strange quark into a nuclear system, a process characterized by low cross sections. Two primary methodologies have been employed: (i) strangeness exchange reactions utilizing kaon beams and (ii) the creation of a $s{\bar s}$ quark pair through high-energy collisions involving beams such as pions, electrons, or ions \cite{Hashimoto2006,Feliciello2015,Gal2016}. Secondly, the sub-nanosecond lifetime of hypernuclei is comparable to the weak decay lifetimes of free hyperons (e.g., 263 ps for the free $\rm \Lambda$~\cite{Zyla2020}). Hypernuclei are produced in their ground states or bound excited states, the latter typically decaying to the ground state via prompt electromagnetic transitions. Light hypernuclei in the ground state predominantly undergo mesonic weak decay, e.g., $_\Lambda^AZ \rightarrow \pi^- + {}^A(Z+1)$, while heavier hypernuclei favor non-mesonic weak decay, a process exclusively occurring within the nucleus \cite{Dubach1996}. Consequently, measurements must be conducted at or proximal to the production site, often resulting in high detector rates and substantial background limitations.\\

    \noindent While ultra-relativistic heavy-ion collision experiments are restricted to the production of hypernuclei up to $\sim$5 baryons \cite{Andronic2018}, ion-ion collisions in the range of a few GeV/nucleon incident energy hold the promise of reaching unexplored regions of the hypernuclear landscape, in particular with the future use of relativistic radioactive beams in inverse kinematics \cite{Sun2018,Saito2021,Aumann2024}. The production of $\rm \Lambda$ hyperons from nucleon-nucleon collisions is reached at $\sim$ 1.6 GeV/nucleon incident energy in the laboratory. The in-flight production of light hypernuclei from ion-ion collisions was pioneered by the HypHI Phase 0 experiment \cite{Saito2016}, which benefited from the large acceptance ALADIN magnet, not in use anymore, at GSI-FAIR. In this first experiment, a $^6$Li beam at 2 GeV/nucleon impinging onto a $^{12}$C target was used to produce and study particularly $^3_{\rm \Lambda}$H and $^4_{\rm \Lambda}$H. The invariant mass of these light hypernuclei was extracted by measuring the pion and light fragment from their two-body weak decay. Signal extraction and optimization of the signal-to-background ratio in the invariant mass spectrum were achieved through coincidence and vertex reconstruction, with a selection criterion requiring the decay vertex to be located a few centimeters downstream of the production target. In this setup, an invariant mass resolution of $\sim (5\pm 1)~\text{MeV}/c^2$ and an efficiency and acceptance of 0.6\% were obtained \cite{Saito2016,Rappold2015}. The lifetimes of these hypernuclei, their production cross section, and a signal of a potential existence of a ${\rm \Lambda} nn$ system were obtained from this measurement.
    \begin{figure}[h!]
        \centering
        \includegraphics[trim=0cm 7cm 0cm 7cm,clip,width=\textwidth]{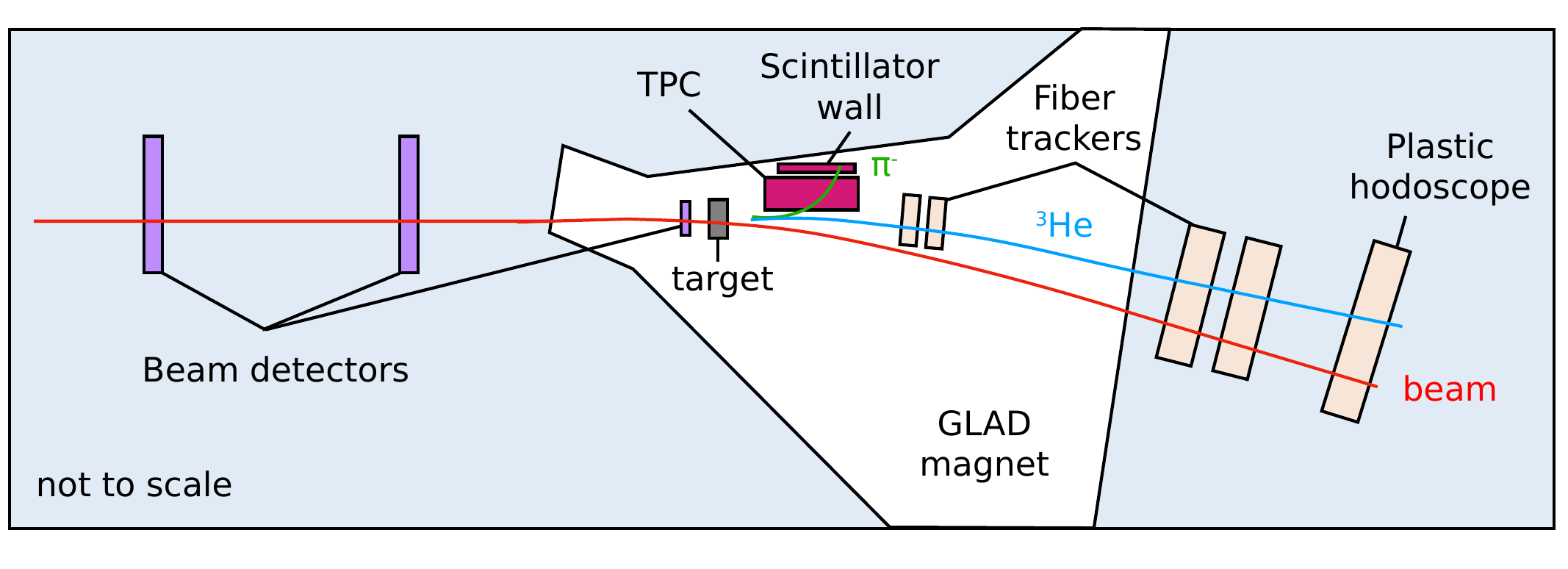}
        \caption{Sketch of the R$^3$B experimental setup including the HYDRA tracker inside the GLAD large-acceptance dipole magnet.}
        \label{fig:r3b}
    \end{figure}
    
    \noindent A long-term program, HYDRA (HYpernuclei Decay at R$^3$B Apparatus), has been proposed within the R$^3$B experimental hall at the FAIR facility to investigate hypernuclei production and properties via relativistic ion collisions. This program utilizes the large acceptance of the R³B GLAD dipole magnet, employing an analogous methodology. Aiming for an improved invariant-mass resolution of $\sim 1.5~\text{MeV}/c^2$ following the mesonic weak decay of the in-flight produced hypernucleus and allowing for high luminosity, a setup inside the bore of the magnet was designed with a dedicated pion tracker, as illustrated in \figref{fig:r3b} and introduced in \cite{Velardita2023}.  
    \\
    
    \noindent In the following, we present the new HYDRA pion tracker to be located inside the GLAD large acceptance dipole magnet of R$^3$B at GSI-FAIR. The pion tracker consists of a hybrid-amplification time projection chamber and a scintillator wall for trigger and timing purposes. We report on the design and construction of both components as well as the first characterization from cosmic ray, source, and laser measurements. 

\section{Requirements}\label{sec:req}
    \noindent The production of hypernuclei from ion-ion collisions accounts approximately for 10$^{-8}$-10$^{-6}$ of the reaction cross section. Several potential background channels lead to the same final state of a $\pi^-$ and a recoil fragment in coincidence, imposing strong constraints on invariant-mass resolution and kinematical event selection for a satisfactory signal over background ratio. Hypernuclei production cross sections of $\mathcal{O}(1~\text{\textmu}\text{b})$ require, in addition, high beam intensities (of the order of 10$^6$ particles per second (pps) or higher) to allow spectroscopy experiments in a reasonable beam time. These general observations lead to the following requirements for the HYDRA pion tracker. Considering these factors that are detailed below, and taking into account that a time projection chamber (TPC) is the most cost-effective solution to be used as a tracker with large acceptance, a TPC has been considered for pion tracking.  \\

    \noindent {\bf Spatial resolution} The invariant-mass resolution is estimated by considering the momentum and angular resolution of the mesonic decayed $\pi^-$ and the fragment. To achieve an invariant-mass resolution of 1.5~MeV/$c^2$ while considering the capabilities of the R$^3$B setup for the fragment tracking, a momentum resolution of 1\% ($\sigma$) for the pion is required, assuming a homogeneous magnetic field of 2~T. The momentum resolution of a TPC arises from two main components: multiple scattering in the active region of the detector \cite{Workman2022} and the intrinsic position resolution of the TPC \cite{Gluckstern1963}. In the targeted experiments at R$^3$B, the momentum of the decay product pion ranges from 200 to 800~MeV/c. In this range, the intrinsic position resolution of the TPC dominates the momentum resolution. At a first order, confirmed by realistic Monte-Carlo simulations, the momentum resolution $\delta p /p$ of the TPC is connected to its intrinsic position resolution $\delta x$ by $\delta p/p \sim \delta x / S$, where $S$ is the sagitta of the projected trajectory on the detection plane of the TPC. For HYDRA, a $\delta p /p=1 \%$ momentum resolution corresponds to a requirement of $\delta x=300 \, \text{\textmu}\text{m}$ position resolution assuming ten equidistant points on the projected track.\\

    \noindent {\bf Vertex reconstruction} The main potential source of background comes from a $\pi^-$ from the decay of a free $\rm \Lambda$ hyperon, a kaon, or a heavier hypernucleus in coincidence with a residue $^A(Z+1)$ nucleus produced from the fragmentation of the projectile. The rejection of these background events mostly relies on the reconstruction of the decay vertex. Realistic Monte-Carlo simulations show that a decay-vertex position resolution better than 10~mm leads to an acceptable signal-over-background ratio of 3. This requirement calls for a minimized radiation length along the pion trajectory, justifying the choice of a gas TPC, as well as the pion and fragment tracking close to the decay vertex.\\

    \noindent {\bf Efficiency and acceptance} Due to the targeted high beam intensity, we designed a geometry as illustrated in \figref{fig:r3b} where the unreacted beam does not pass through the tracking detectors. As the weak-decay vertices distribute over several tens of centimeters downstream of the target, a spatially extended pion tracker is required to achieve adequate acceptance and detection efficiency. Within the spatial constraints of the bore of the GLAD magnet, a 30-cm long active region for tracking was chosen that provides a combined acceptance and detection efficiency of 25\%. For such a large tracker, a time-projection chamber is the most cost-effective solution to achieve the above requirements. \\

    \noindent {\bf Ion back-flow (IBF)} A main technical challenge to the choice of a TPC is the ion back-flow, characterized by the drift of amplification-generated ions back into the drift volume. This phenomenon can adversely affect the uniformity of the drift field, resulting in a degradation of spatial resolution. At the optimal relative position of the production target and the TPC inside GLAD, realistic simulations give 200~kHz of charged particles, mostly protons from target fragmentation, traversing the drift volume of the TPC for a luminosity $\mathcal{L}=7 \cdot 10^{29}~\text{cm}^{-2}~\text{s}^{-1}$, a typical value for target experiments with HYDRA. This rate results in a significant space charge density within the TPC's active volume due to the ion back-flow. To mitigate electric field distortions caused by the space charge, an ion back-flow of less than 1\% is necessary for an amplification gain of 4000 for the amplification region. This requirement corresponds to an average space charge density of $40~\text{fC}~\text{cm}^{-3}$ in the full volume and local maxima reaching $50~\text{fC}~\text{cm}^{-3}$. As a reference, the mean charge density in the ALICE TPC, operating at 50 kHz Pb–Pb collision rates during LHC Run 3, is expected to exceed $100~\text{fC}~\text{cm}^{-3}$ with a drift field of 400 V/cm~\cite{Ball2014}.

\section{Time Projection Chamber}

    \noindent The HYDRA TPC is housed within a cuboid gas vessel, filled with a counting gas at atmospheric pressure. As charged particles pass through the active volume, they ionize the gas, generating primary ionization electrons that drift along the field toward the amplification stage. The latter comprises a Gas Electron Multiplier (GEM) and a Micromegas structure. The amplified signals are then induced on the pad plane, where their arrival positions are recorded. By utilizing an external reference signal from the scintillator wall, the drift time of the ionization electrons can be determined, enabling reconstruction of the charged particle's trajectory in three dimensions.\\

    \begin{figure*}[h!]
        \centering
        \includegraphics[width=\textwidth]{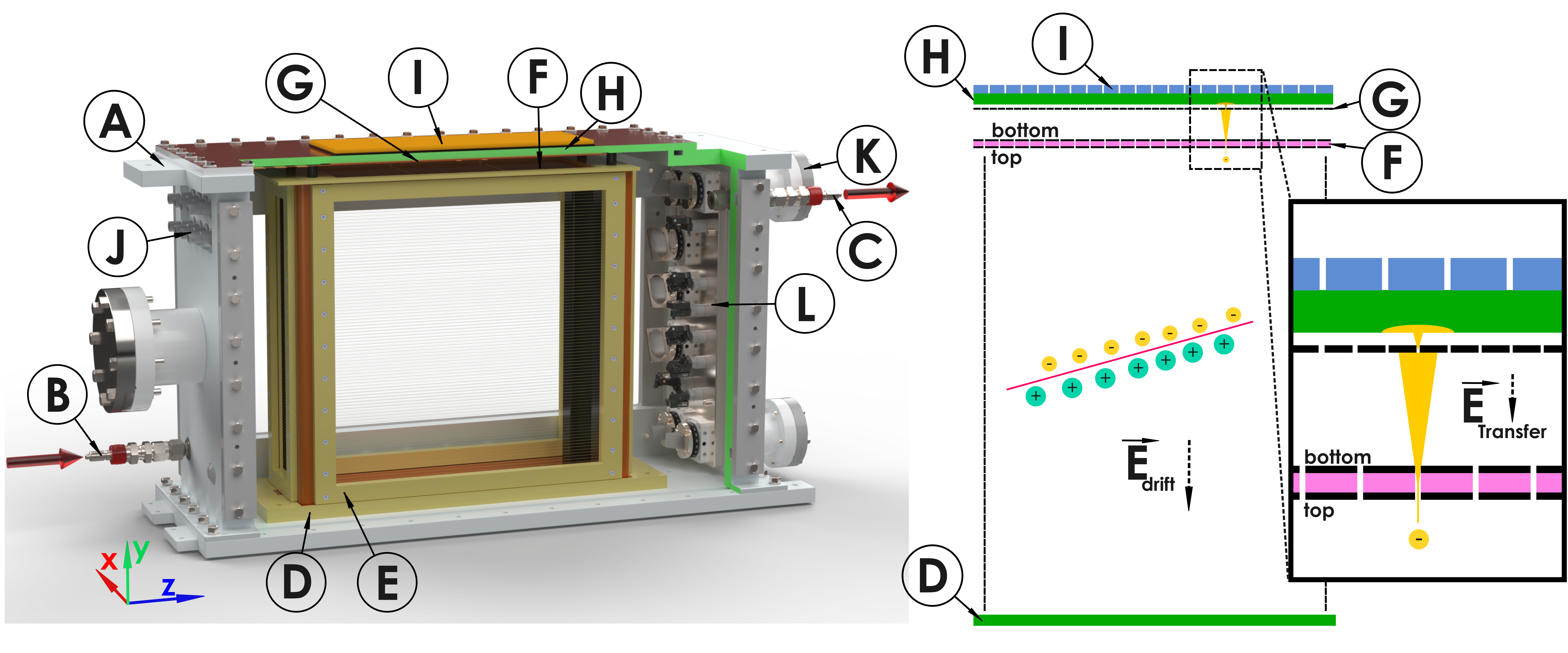}
        \caption{Sectional view of the HYDRA time projection chamber (left), together with drift and amplification regions (right).
        The following components of the TPC are identified in the figure: gas vessel (A) with gas inlet (B) and outlet (C), drift cathode PCB (D), field cage (E), GEM foil (F), Micromegas (G), diamond-like carbon layer (H) on top of the metal core pad plane (I), high-voltage feedthroughs (J), laser port (K) and mirror assembly (L).}
        \label{fig:outline}
    \end{figure*}
    
    \noindent The TPC’s drift region is enclosed within a cuboid field cage, with the drift field generated by the drift cathode and GEM electrode potentials, together with a precision resistor chain positioned along the field cage corners. The active volume of the TPC is determined by the active area of the readout anode (the pad plane). A summary of the design parameters and dimensions is provided in Table \ref{tab:dimensions}, while \figref{fig:outline} presents a schematic overview of the TPC.

    \begin{table}[h!]
        \caption{Dimensions and parameters of the HYDRA time projection chamber. The mesh density is given in units of lines per inch (LPI).}
        \centering\footnotesize 
        \begin{tabular}{ l l c c }
            \hline
            & Parameter & Value & Unit \\
            \hline
            {\bf Vessel} & length ($Z$) & 555 & mm \\
            & width ($X$) & 229 & mm \\
            & height ($Y$) & 400 & mm \\
            & mass & 30 & kg \\
           {\bf  Drift region} & drift gap ($Y$) & 300 & mm \\
            & cathode area ($X\times Z$) & 113 $\times$ 334 & mm$^2$ \\
            {\bf  GEM}& area ($X\times Z$) & $96\times300$ & mm$^2$ \\
            & total thickness & 60 & $\upmu$m \\
            & Cu thickness ($\times 2$) & 5 & $\upmu$m \\
            & outer hole $\diameter$ & 70 $\pm$ 5 & $\upmu$m \\
            & inner hole $\diameter$ & 50 $\pm$ 5 & $\upmu$m \\
            & hole pitch & 140 & $\upmu$m \\
            & transfer gap & 5 & mm \\
            {\bf  Micromegas}& mesh density & 400 & LPI \\
            & gap & 128 & $\upmu$m \\
            & DLC resistance & 1.2 - 1.5 & M$\rm{\Omega}/\square$ \\
            {\bf  Pad plane} & active area ($X\times Z$) & $87.9\times255.9$ & mm$^2$ \\
            & pad size & $1.9 \times 1.9$ & mm$^2$ \\
            & inter-pad distance & 0.1 & mm\\
            {\bf  Field-cage}& Cu-Be wire $\diameter$ & 75 & $\upmu$m \\
            & wire tension & 80 & cN/wire\\
            & wire spacing & 3 & mm\\
            &inter-plane gap & 4& mm\\
            \hline
        \end{tabular}
        \label{tab:dimensions}
    \end{table}
        
    \subsection{Gas Vessel and Gas System}
        \noindent The gas vessel with an approximate internal volume of 42~dm$^3$ is made of non-ferromagnetic EN AW-5082 aluminum alloy. Two walls of the vessel, parallel to the $YZ$-plane, were machined to accommodate thin windows, minimizing energy loss and angular straggling for particles passing through the TPC. The windows contain 12~$\upmu$m aluminized Kapton foils glued onto the aluminum frames. One side wall of the vessel is equipped with SHV-5 connectors and a single Lemo high-voltage (HV) connector (RAD.00.113.CTM), rated for a maximum voltage of 10~kV. QC-4 Swagelok connectors are used to regulate gas flow inside the vessel, maintaining a stable flow rate, adjustable within the range of 5 to 20~L/h. The vessel and the gas system are gas-tight. The O$_2$ impurity was surveilled and kept below 30~ppm (parts per million) during the measurements.\\
    
        \noindent The TPC is equipped with an open-loop gas system, without the regulation of overpressure, which reaches typically up to 10~mbar depending on the gas flow. To meet the performance requirements described in \secref{sec:req}, the counting gas should provide a high drift velocity and moderate diffusion properties. As a potential candidate, the so-called T2K mixture of \TtwoK \cite{Koch2017} is considered. This gas mixture exhibits the electron drift velocity of $\sim$8\,cm/$\upmu$s at a drift field of 275~V/cm. At this field strength, the maximum applied cathode potential can be limited to below 10~kV. This limitation reduces the risk of electrical discharge within the TPC, as the shortest distance between the cathode electrode and the grounded vessel wall is 15~mm. Furthermore, the maximum electron drift velocity minimizes the sensitivity of the drift velocity to minor fluctuations in the electric field. For the commissioning of the TPC, presented in this work, an \ArCOtwo mixture is used with a drift field of 220~V/cm. Table \ref{tab:gas} compares the main properties of both mixtures.

        \begin{table*}[h!]
        \caption{Properties of \TtwoK gas mixture considered for the operation of the HYDRA TPC and \ArCOtwo used for commissioning of the detector.  The drift velocity and diffusion coefficients are evaluated at the respective drift field values and zero magnetic field using Magboltz~\cite{SBiagi2000}. The effective ionization potential values for \TtwoK and \ArCOtwo are taken from~\cite{Atti2023} and~\cite{Gasik2023}, respectively.}
        \centering\footnotesize
        \begin{tabular}{ l c c }
            \hline
            & \TtwoK & \ArCOtwo \\
            \hline
            {\bf Drift field [V/cm]} & 275 &  220\\
            {\bf  Drift velocity [cm/$\mathbf {\upmu}$m]} & 7.82 & 1.56 \\
            {\bf  Transverse diffusion [$\mathbf{ \sqrt{\mathrm{\mathbf {cm}}}}$]} & 0.0323 & 0.0222 \\
            {\bf  Longitudinal diffusion [$\sqrt{\mathrm{\mathbf{cm}}}$]} & 0.0215 & 0.0250 \\
            {\bf  Eff. ionization potential [eV]} & 26.8 & 28.8 \\  
            \hline
        \end{tabular}
        \label{tab:gas}
        \end{table*}

    \subsection{Field Cage and Cathode}
\begin{figure}[h!]
        \centering
        \includegraphics[trim={7cm 0cm 7cm 0cm},clip,width=0.75\textwidth]{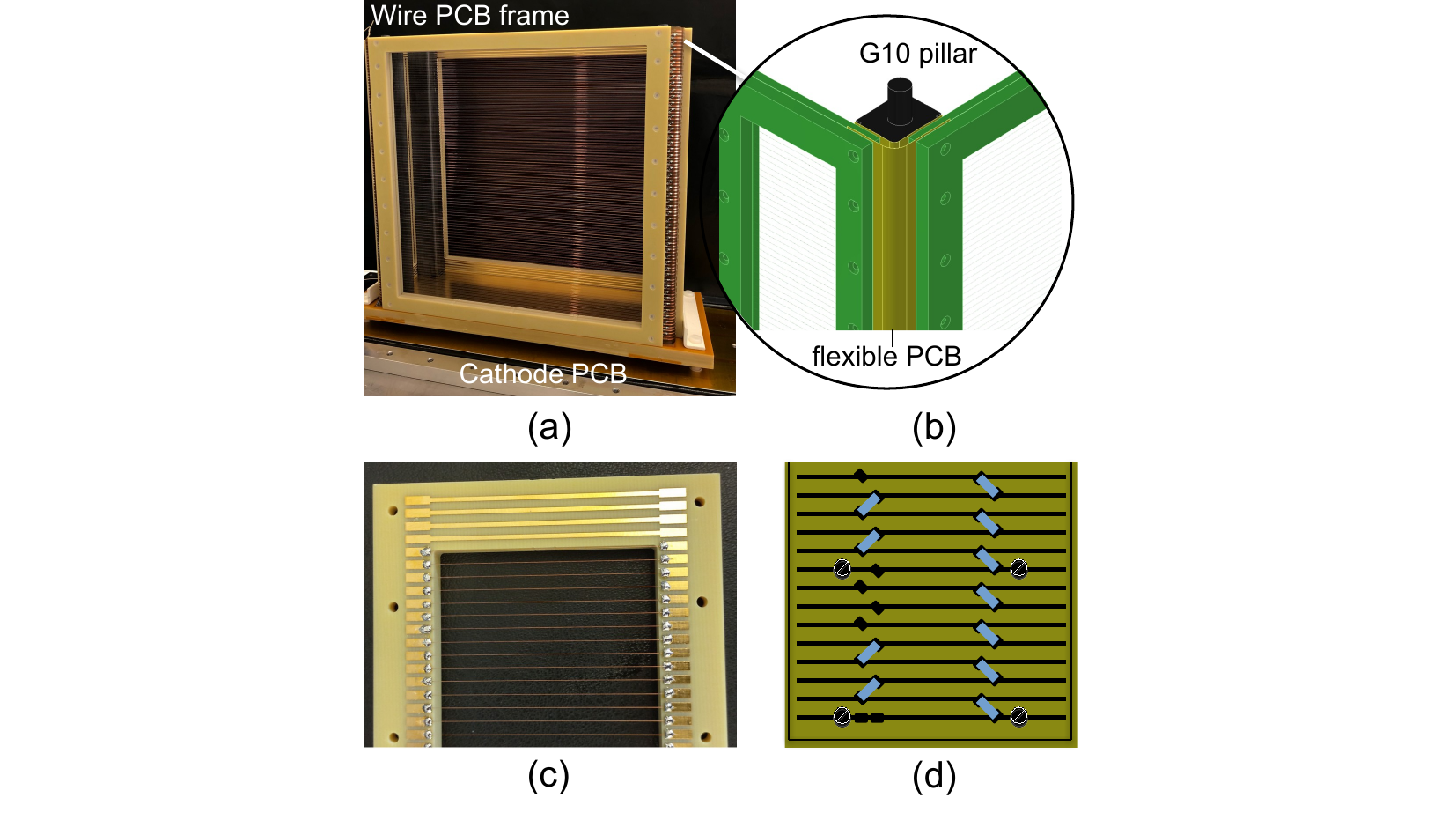}
        \caption{Details of the HYDRA field cage: wire planes soldered on FR4 frames (a,c), electrical connection of wire planes by flexible PCBs (b) mounted onto G10 pillars. The potential gradient is imposed by surface-mounted resistors (positions shown in blue in panel (d)) between Ni/Au-platted Cu tracks.}
        \label{fig:FC}
        \end{figure}

        \noindent The wired field cage is based on the design of the CAT time projection chamber \cite{Ota2015}. Details of the field cage are illustrated in \figref{fig:FC}. It has a geometrical transparency of 94.8\%. The field cage is composed of two layers of wires fixed onto frames made of FR4 glass epoxy frames (EMC 370-5). Each layer consists of two wide and two narrow frames. They are separated by a 4~mm gap given by the thickness of the frames. Equipotential wires of different frames are electrically connected by flexible PCBs (panel (b) of \figref{fig:FC}). 
        All field cage frames are fixed with nylon? screws to G10 pillars, which are inserted into the drift cathode PCB. The drift electrode comprises a 35~\textmu m thick copper layer, coated with nickel and gold to improve corrosion resistance and electrical conductivity, and deposited onto a 13~mm thick FR4 substrate.\\
    
        \noindent The wires have a diameter of 75~$\upmu$m and are made of copper-beryllium alloy 25 (C17200) with 98\% Cu and 2\% Be, characterized by high conductivity, hardness, high corrosion resistance, and excellent spring properties with a density of 8.3~g/cm$^3$. They were tempered, age-hardened, and cold-drawn in advance, optimizing their ductility and enabling them to hold their straight form by tension. In each plane, they are spaced by 3~mm. The average tension imposed on the wires is approximately 80~cN per wire. To ensure a homogeneous drift field near the cathode and the amplification region, the field cage frames incorporate on each side four 1~mm wide strips with a 3~mm pitch matching the inter-wire spacing of the field cage. The flexible PCBs are composed of the same strip pattern. The inner layer of wires is shifted by 1.5~mm vertically from the outer layer to avoid punch-through fields inside the drift region and guarantee a homogeneous drift field also close to the wires, as detailed further. The field cage frames and the flexible PCBs were produced at the EP-DT-DD Micropattern Structures Laboratory of CERN, while the wiring process was carried out at the GSI Detector Laboratory using a winding machine.\\

        \begin{figure}[h!]
        \centering
        \includegraphics[width=0.75\textwidth]{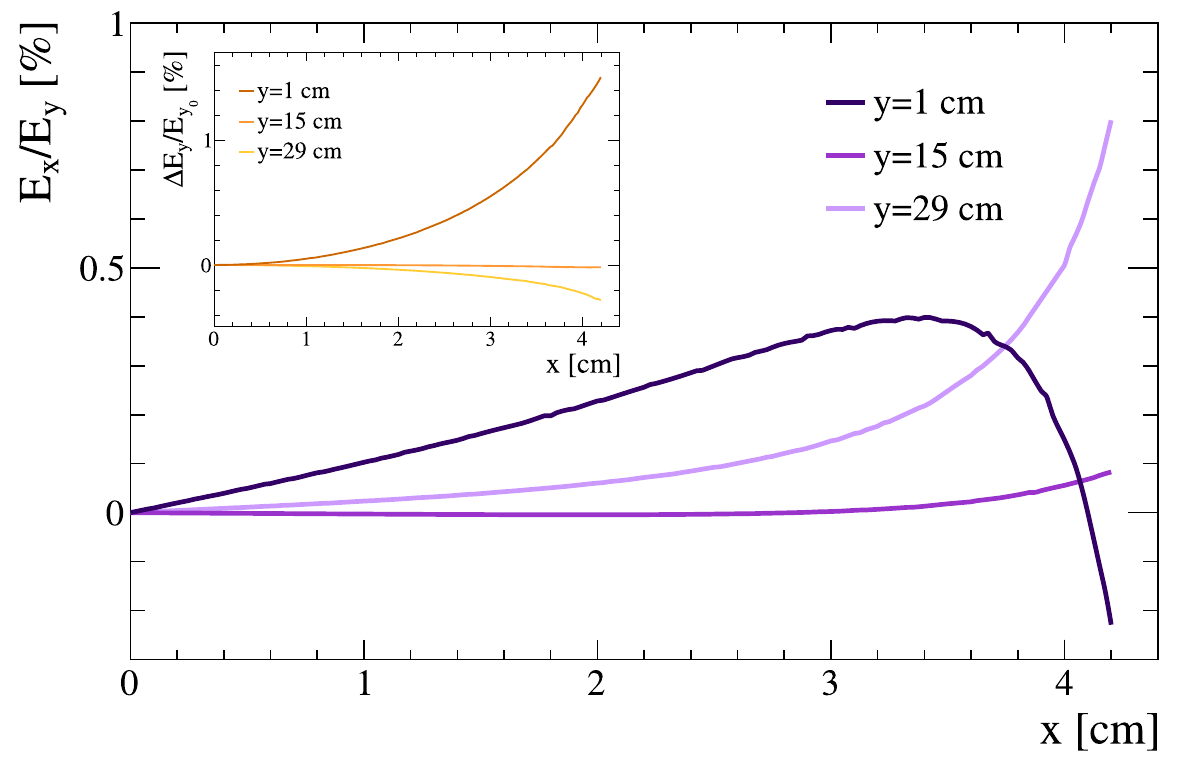}
        \caption{Simulated ratio of transverse component (parallel to the cathode) $\rm{E_x}$ to the longitudinal component (perpendicular to the cathode) $\rm{E_y}$ of the drift field. The $x$-axis ranges from 0, the center of the cathode, to 4.2~cm, the edge of the active area. The different curves correspond to a vertical $y$-position at the center of the drift field ($y=15$~cm), 1 cm from the cathode ($y=1$~cm), and 1~cm from the GEM ($y=29$~cm). The inset shows the relative deviation of the vertical drift field component from that at $x=0$.}
        \label{fig:field-non-uniform}
        \end{figure}
    
        \noindent The flexible PCBs are made of 50 $\upmu$m Kapton. The tracks made of 17~$\upmu$m thick, passivated copper with Ni/Au plating, include the surface-mounted (SMD) resistor solder pads (footprints). The potentials of the field cage wires and strips are defined with 1 M$\rm{\Omega}$ resistors soldered between the tracks of the flexible PCBs. The first strips of the inner and outer layers are connected to the cathode through 0.5~M${\rm \Omega}$ and 0.75 M${\rm \Omega}$ axial-lead resistors, respectively.
        The last strips of the outer layer are connected via a 0.25~M${\rm \Omega}$ axial-lead resistor to the last strips of the inner layer. The latter are connected to an external HV feedthrough via a 0.25~M${\rm \Omega}$ axial-lead resistor. The potential of the last strip can be adjusted to match the potential of the top GEM electrode. All the aforementioned resistors were chosen with a tolerance of 1\% and a power rating of 0.25~W.\\ 
    
        \noindent The electric field within the drift volume was simulated using finite element method software, namely Gmsh \cite{Geuzaine2009} and ElmerFEM \cite{Malinen2013}. The field non-uniformity, defined as the ratio of transverse electric field component ($E_x$) to the longitudinal component ($E_y$) as illustrated in \figref{fig:field-non-uniform}, reached a maximum of 1.5\% at the periphery of the active volume, while within the central region, the non-uniformity remained below 0.5\%. In the more central region, the distortion is generally less than 0.5\%. For the ratio $\Delta E_y/E_{y(x=0)}$ (with $\Delta E_y = E_y - E_{y(x=0)}$), it is less than 1.6\%. The non-uniformity of the drift field is one primary contribution to the spatial resolution by displacing the drift electrons. While the calculated non-uniformity is deemed adequate for achieving the required spatial resolution, a quantitative evaluation of its impact on electron drift necessitates the following simulations.\\
        
\begin{figure}[h!]
        \centering
        \includegraphics[width=0.75\textwidth]{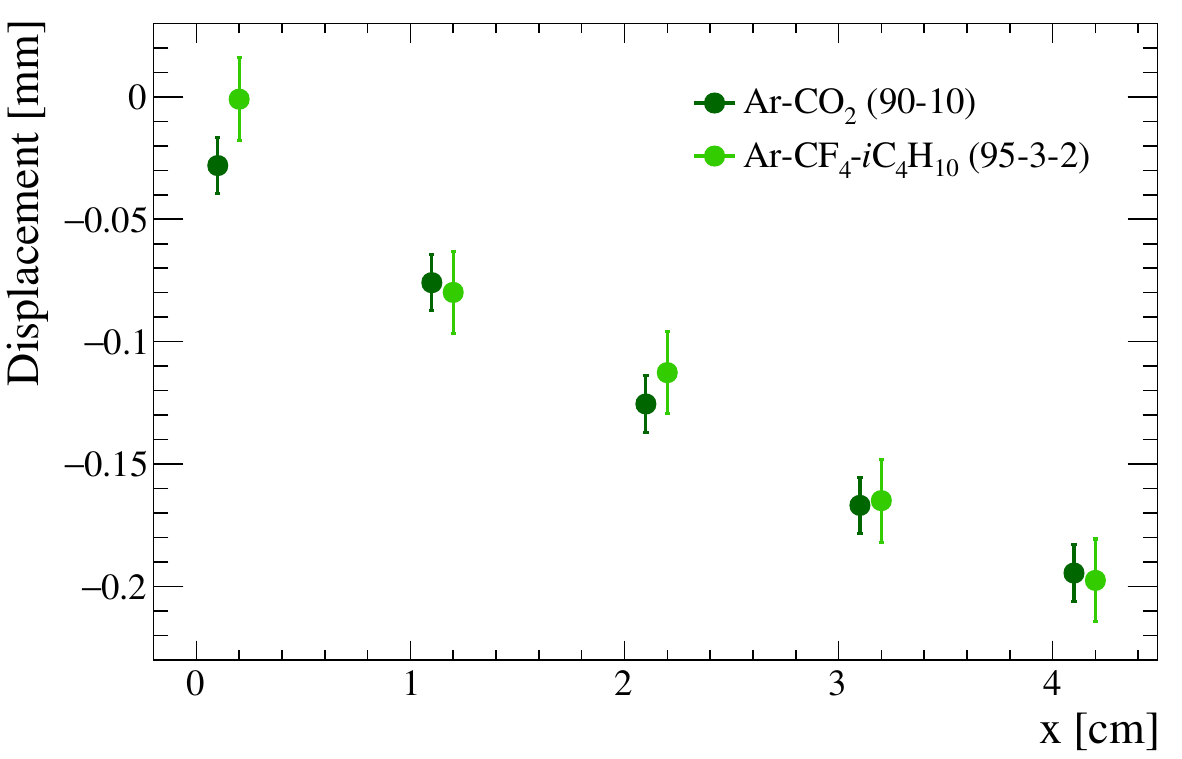}
        \caption{Simulated displacement of electron clouds as a function of the initial position of drift electronics. Simulations were performed with two different gas mixtures: \TtwoK and \ArCOtwo. The dark green points are shifted by 0.1~cm along the x-axis for better visibility. }
        \label{fig:electron-drift}
        \end{figure}

        \noindent Using the calculated drift field, simulations of electron drift were performed with Garfield++ \cite{Schindler2023}. Electrons were released 25~cm from the anode along the drift direction and at different perpendicular positions from the center of the active volume to the periphery and drifted towards the anode along the field lines. Upon arrival at the anode, the electrons can deviate from their initial perpendicular positions, either due to diffusion or caused by non-uniformities of the drift field and/or non-zero magnetic field perpendicular to the drift field. Assuming no magnetic field and considering the realistic electric field in the simulation, a maximum mean displacement of around 200~$\upmu$m was obtained at the periphery of the active volume, as illustrated in \figref{fig:electron-drift}, while in the central region, the displacement is well below 200~$\upmu$m. It is therefore concluded the the non-uniformity of the drift field meets the requirement of the spatial resolution for the TPC. In addition, the effect of inhomogeneities in the GLAD magnetic field on the tracking resolution, not addressed here, is to be considered during the data analysis.

    \subsection{Amplification Region}\label{sub:amp}
        \noindent The amplification region is a two-stage hybrid detector incorporating a GEM \cite{Sauli1997} and a resistive Micromegas \cite{Giomataris1996,Atti2022} positioned at the end of the drift region, as illustrated in \figref{fig:outline}. This hybrid combination is expected to achieve an overall IBF rate below 1\% \cite{Zhang2017}. All components were manufactured at the EP-DT-DD Micropattern Structures Laboratory of CERN.\\
    
\begin{figure}[h!]
        \centering
        \includegraphics[trim={0 1cm 0 1cm},clip,width=0.75\textwidth]{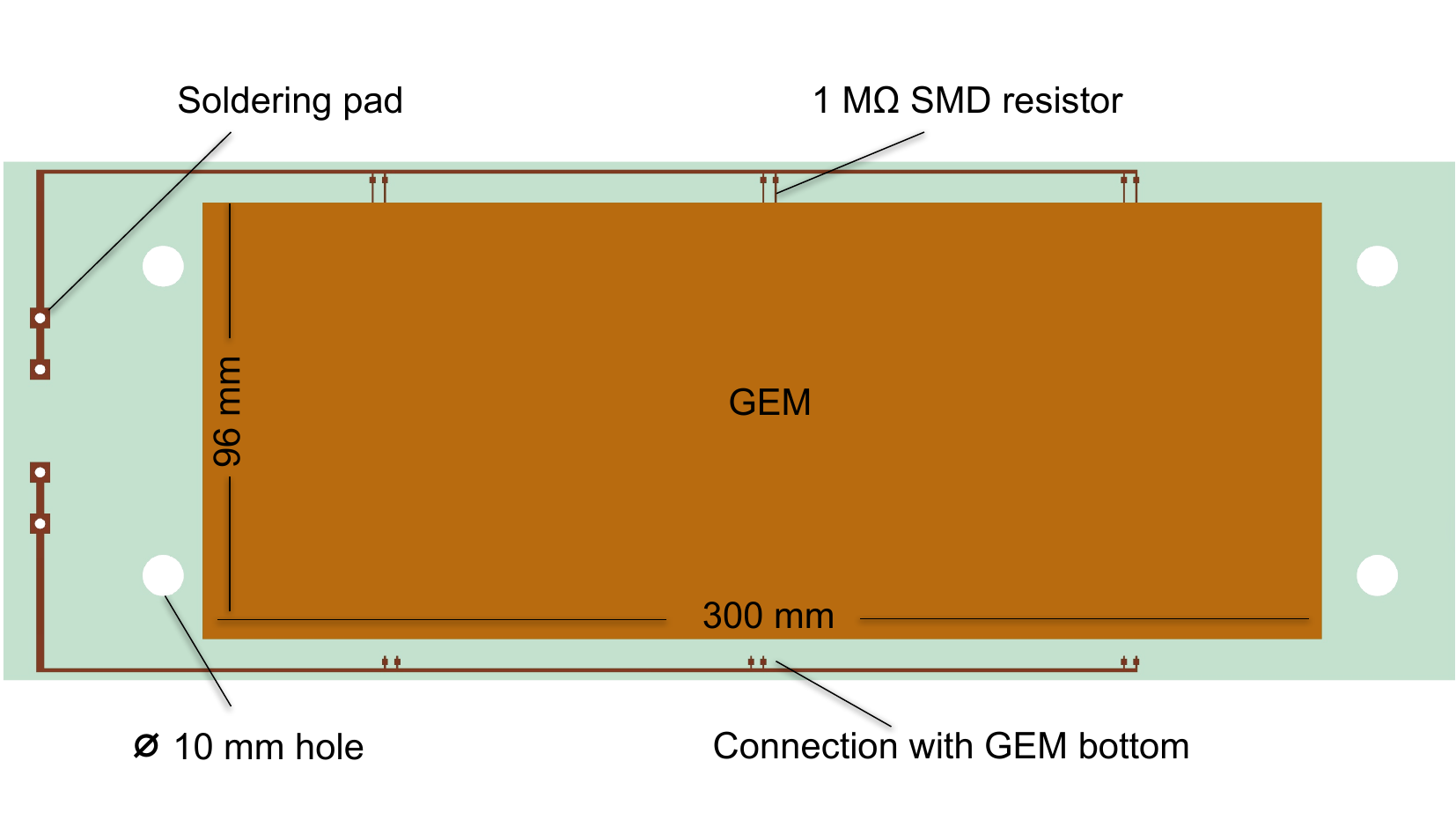}
        \caption{Schematic of the top side of the GEM foil. The orange rectangle outlines the total active area; the three internal GEM segments are not depicted. Brown tracks indicate the high-voltage path, and the light green area marks the frame gluing region.}
        \label{fig:gem}
        \end{figure}

        \noindent The initial amplification stage is a 300 $\times$ 96 mm$^2$ GEM foil (see \figref{fig:gem}) segmented into three parts on both sides. The GEM foil is stretched and glued onto a 3~mm thick frame made of FR4 glass-reinforced epoxy laminate (EMC 370-5). Both sides of the GEM on each segment are connected to soldering pads via 1M$\rm \Omega$ surface-mounted resistors. Radiation-resistant wires with PEEK insulation (Allectra 311-PEEKM-035-10M) are soldered on these pads, connecting both sides of the GEM separately to SHV-5 connectors.\\
    
        \noindent The GEM is positioned on top of the field cage and fixed to the supporting G10 pillars using flat-headed screws. The top side of the GEM foil, facing the cathode (see \figref{fig:outline}), contributes to the uniformity of the drift field in conjunction with the cathode, the wires, and the strips of the field cage. Furthermore, the GEM foil extends beyond the edges of the active area, defined by the dimensions of the pad plane (see Table \ref{tab:dimensions}), ensuring a uniform drift field throughout the active volume.\\

        \noindent The second stage of the amplification region is a resistive Micromegas. A gap of five millimeters separates the GEM and the Micromegas, as illustrated in \figref{fig:outline}. This gap can be easily modified within the range of 5 - 11 mm by adding precisely machined spacers. Within this gap, a transfer field is applied to extract electrons from the GEM and transport them to the Micromegas. Simulations of electron drift in the transfer gap of 5 mm exhibit a maximum field distortion of $E_x/E_y$ = 1.2\% within the active area, leading to a maximum electron transverse displacement of less than 40 $\upmu$m after drifting through the transfer gap.\\
        
        \noindent The Micromegas mesh is stretched over the readout anode and supported with 128~$\upmu$m thick pillars made of photopolymer solder mask material (DuPont\textsuperscript{TM} Pyralux\textsuperscript{\textregistered} PC 1000). The readout pads are superimposed by a single diamond-like carbon (DLC) layer, deposited on an insulating film. The measured surface resistivity of the DLC layer varies between 1.2 and 1.5~M$\Omega$/$\square$. The mesh and the DLC layer can be biased with separate HV channels.\\

        \begin{figure}[h!]
        \centering
        \includegraphics[trim={3cm 0cm 5cm 0cm},clip,width=\textwidth]{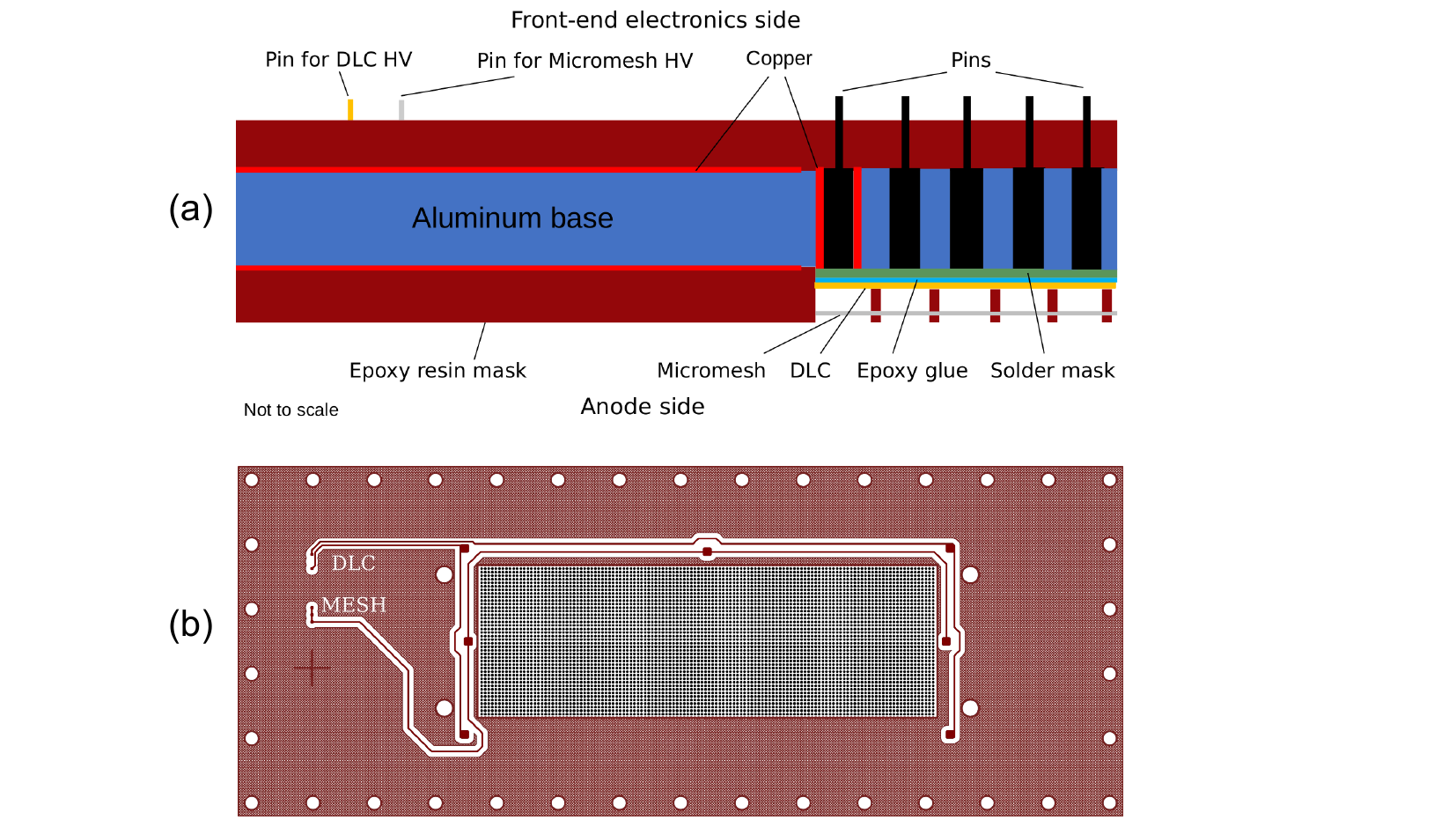}
        \caption{(a) Scheme of the Micromegas-embedded metal core pad plane. The red lines are grounding areas. The outermost rows of pins are independently grounded. (b) Design of the corresponding anode side.}
        \label{fig:mcpp}
        \end{figure}
    
        \noindent The metal core pad plane (MCPP) \cite{Giovinazzo2018} has dimensions of 497.8 $\times$ $197\times7$~mm$^3$ and incorporates the resistive Micromegas, SAMTEC MMTM-144-05-L-D-100 pin connectors serving as readout pads, and a bulk aluminum plate base, as depicted in \figref{fig:mcpp}. The pad plane consists of $44 \times 128$ square pads, each covering an area of $1.9 \times 1.9$ mm$^2$, separated by a gap of 0.1 mm. In total, 5632 pads occupy an area of 224.94~cm$^2$, which is the active area of the TPC. The primary consideration for employing the MCPP structure is its mechanical stability, which enables operations with pressures lower or higher than the atmospheric pressure, if required.\\
        
        \noindent The entire aluminum substrate is coated with copper on both sides of the pad plane, as illustrated in panel (a) of \figref{fig:mcpp}, and is grounded once mounted onto the TPC vessel. There are two rows of pads (the first and last rows in panel (b) of \figref{fig:mcpp}) connected to a dedicated copper area and subsequently shielding pins. During operation, these shielding pins are connected to the vessel, ensuring that the whole TPC and the electronics share the same ground.\\

        \noindent The implementation of this hybrid Micromegas-GEM amplification is intended to reach an IBF below 1\%. Previous investigations employing comparable hybrid structures have demonstrated IBFs of less than 0.5\% at a gain of 4000 with the \ArCOtwo gas mixture \cite{Zhang2017} and approximately 0.1\% at a gain of around 5000 with the \TtwoK gas mixture \cite{CEPCTPC_Zhiyang_yuan}. A dedicated characterization of the IBF performance for the HYDRA-TPC is scheduled in the near future.
    
    \subsection{High Voltage Scheme}\label{TPC:HV}
\begin{figure}[h!]
        \centering
        \includegraphics[trim=1.5cm 0cm 1.5cm 0cm,clip,width=0.75\textwidth]{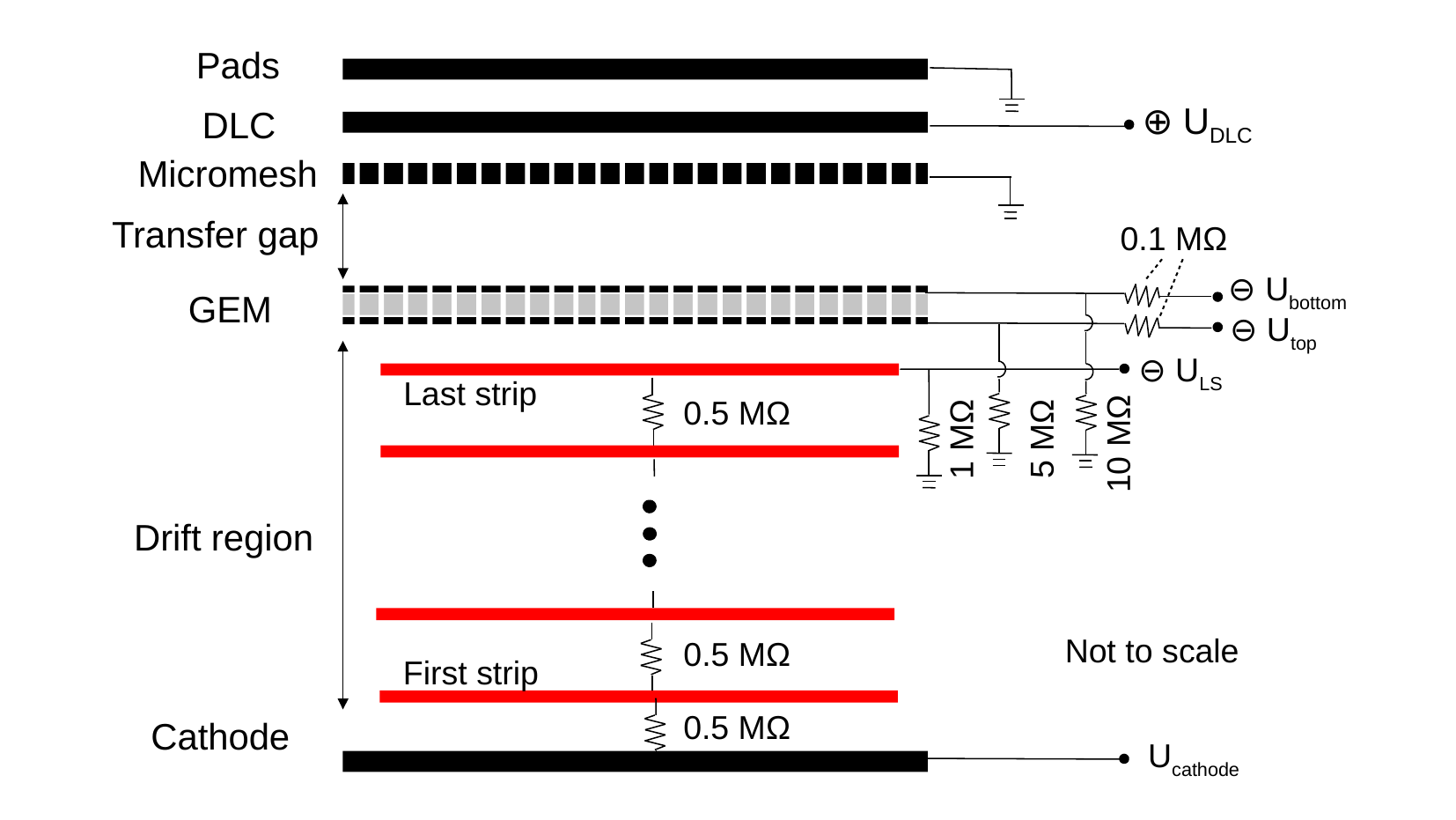}
        \caption{High voltage scheme of the HYDRA-TPC.}
        \label{fig:hv-scheme}
        \end{figure}

        \noindent A high voltage (HV) scheme was designed to provide a guideline for the application of HV to the five electrodes and the operation of the TPC, as shown in \figref{fig:hv-scheme}. Four electrodes, namely the cathode (biased $\rm U_\mathrm{cathode}$), the last strip ($\rm U_\mathrm{LS}$), the top side of the GEM ($\rm U_\mathrm{top}$), and the bottom side of the GEM ($\rm U_\mathrm{bottom}$), are connected to a negatively polarized power supply. The DLC layer is set to a positive potential ($\rm U_\mathrm{DLC}$), generating the amplification field within the Micromegas structure.\\
    
        \noindent The HV potentials of the electrodes are set to guarantee the uniformity of the drift field between the cathode and the GEM and to ensure a total effective gain in an operational range from $2\times10^3$ to $10^4$. An example of voltage settings applied to the electrodes of the TPC during commissioning measurements with the laser system using the \ArCOtwo gas mixture, resulting in a gain of $\sim$$10^4$, is provided in Table \ref{tab:voltages}.\\
    
        \noindent The ramping speed of the HV power supplies is controlled to avoid discharges between any of the electrodes. A safety loop is set up to ensure that when the current limit is reached, the HV on all electrodes is turned down immediately. Resistors to ground for GEM HV channel are chosen to assure safe discharge of the GEM foils after an HV trip. A suitable resistor to ground for the last strip connection is chosen to allow for the sinking of a small current to ground.

        \noindent

        \begin{table}[h!]
        \caption{Typical potentials applied to the electrodes of the TPC, and the resulting drift field ($E_{\mathrm{drift}}$), potential difference across GEM ($\Delta U_{\mathrm{GEM}}$), and transfer field between  GEM and Micromegas ($E_{\mathrm{transfer}}$) during commissioning measurements with the laser system using \ArCOtwo gas mixture.}
        \centering
        \begin{minipage}{0.27\textwidth}
        \centering
        \begin{tabular}{|l|c|}
        \hline
        U$_\mathrm{cathode}$  & -7400\,V\\
        U$_\mathrm{LS}$  & -866\,V \\
        U$_\mathrm{top}$  & -800\,V \\
        U$_\mathrm{bottom}$  & -420\,V \\
        U$_\mathrm{DLC}$  & 400\,V \\
        \hline
        $E_{\mathrm{drift}}$ & 220\,V/cm\\
        $\Delta U_{\mathrm{GEM}}$ & 380\,V\\
        $E_{\mathrm{transfer}}$ & 840\,V/cm\\
        \hline
        \end{tabular}
        \end{minipage}

        \label{tab:voltages}
        \end{table}
    \subsection{Laser system}\label{sec:laser}
        \noindent A dedicated laser system is developed to generate reference tracks inside the TPC, in order to calibrate the electron drift velocity. Furthermore, since the TPC is to be operated inside the GLAD magnet, which presents significant magnetic field spatial non-homogeneity, the laser tracks provide benchmark data to measure the displacement on electron drift trajectories and correct the field non-homogeneity. The concept is based on the STAR \cite{Lebedev2002,Abele2003} and ALICE \cite{Renault2007} laser systems. \\
    
\begin{figure}[h!]
            \centering
            \includegraphics[width=0.5\textwidth]{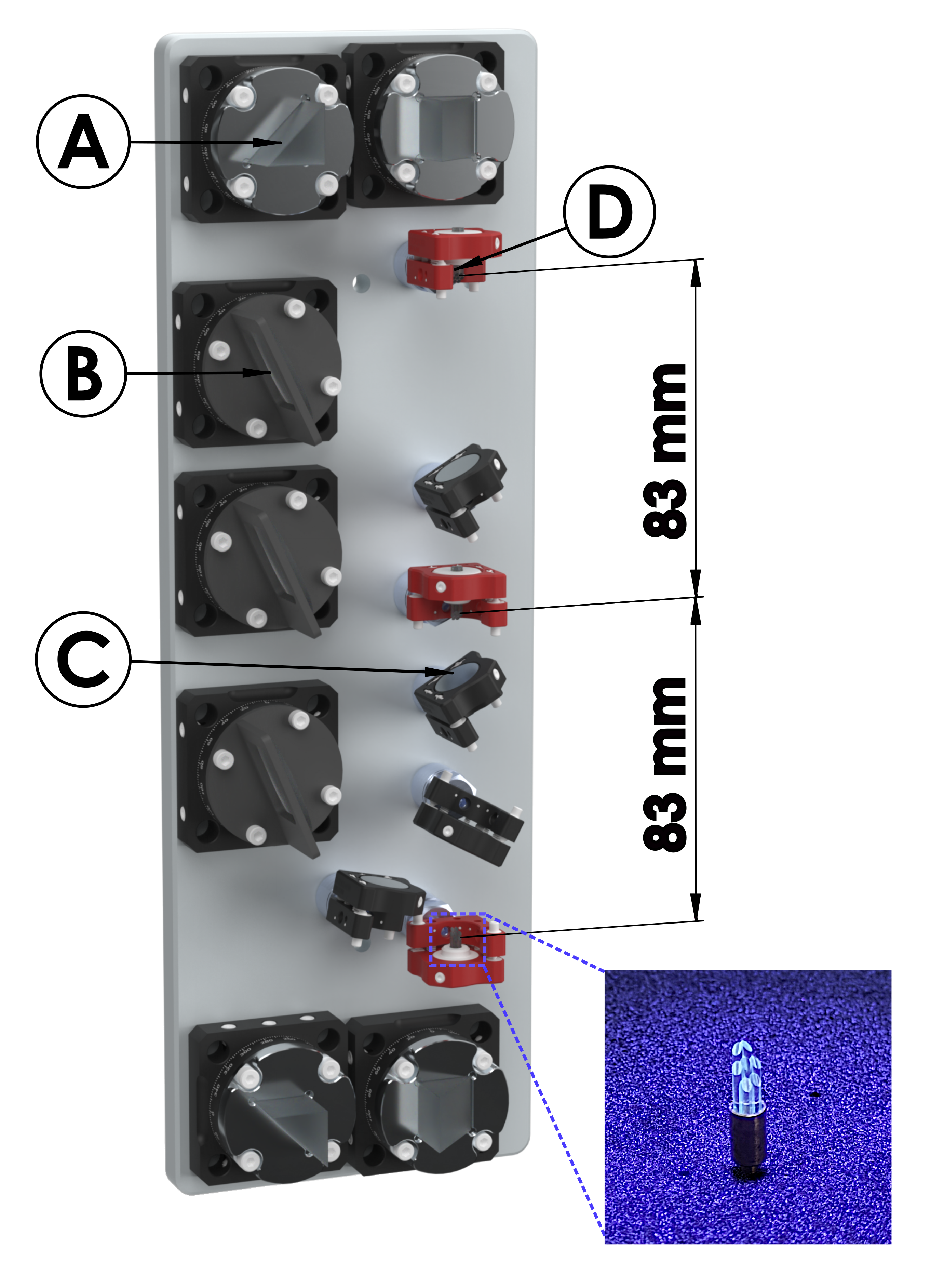}
            \caption{Schematic 3D drawing of the mirror assembly placed inside the TPC vessel: the primary laser is directed into the vessel and reflected by the prisms (A) and subsequently divided into equal intensity beams via beam splitters (B). These sub-beams are then reflected by the mirrors (C) onto micro-mirror bundles (D). A prototype of a micro-mirror bundle provided by A. Lebedev from the STAR collaboration is shown in the insert.}
            \label{fig:laser}
        \end{figure}

        \noindent The system comprises an ultraviolet (UV) laser source and a mirror assembly placed inside the TPC vessel as illustrated in \figref{fig:laser}. The laser source is a 266~nm VironTM diode-pumped Q-switched Nd:YAG laser, with a nominal power output of 9.37~mJ and a maximum pulse repetition rate of 20~Hz. The laser beam exhibits a Gaussian distribution with a mean intensity of 2.48~mJ/mm$^2$. UV lasers emit photons with energies below 4.7~eV (266~nm), which is lower than the ionization energy of most gas atoms and molecules commonly used in TPCs. For instance, the lowest ionization energies for Ar and CO$_2$ are 15.7 eV and 14.4 eV, respectively \cite{W-value}. Still, UV lasers can induce two-photon ionization of organic impurities with ionization potentials of 5-8~eV present in the gas mixture.\\

        \noindent Upon activation, the laser beam is directed through a quartz window mounted on the KF-40 flange into the TPC. A series of reflective mirrors and fused silica beam splitters (BS) of the mirror system, as depicted in \figref{fig:laser}, subsequently split the laser beam into three sub-beams, each 17\% of the primary beam. These sub-beams are then projected onto micro-mirror bundles, each consisting of seven 1-mm diameter micro-mirrors oriented at different angles. The micro-mirrors reflect the laser beams in various directions, generating laser tracks within the active volume of the field cage. These laser tracks produce primary ion-electron clusters through two-photon ionization.

\section{Scintillator wall}
    \subsection{Geometry and mechanics}
\begin{figure}[h!]
        \centering
        \includegraphics[trim=0cm 0cm 0cm 0cm,clip,width=0.75\textwidth]{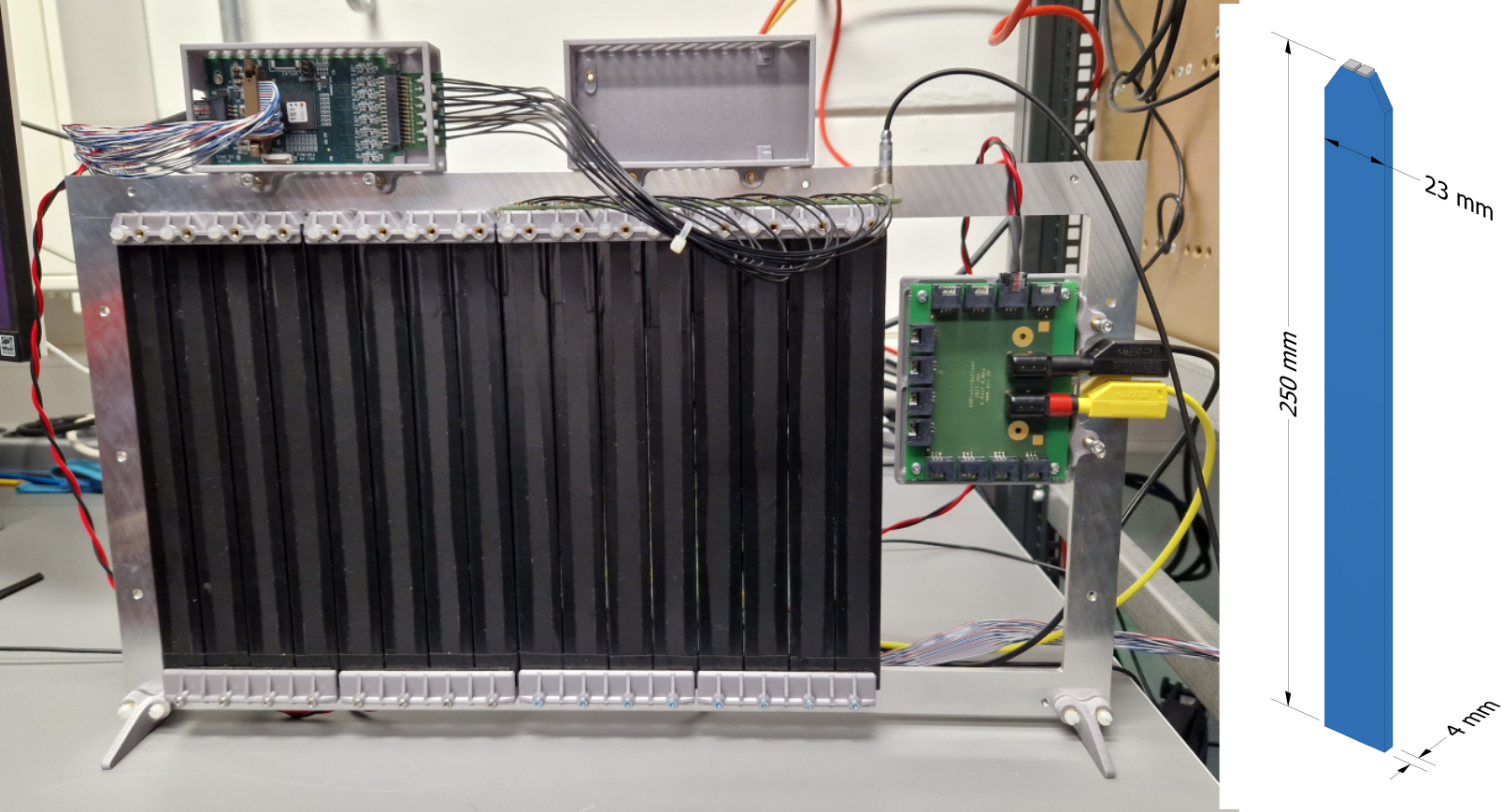}
        \caption{(Left) HYDRA scintillator wall before being wrapped and covered. (Right) sketch of one detection module.}
        \label{fig:pw}
        \end{figure}

        \noindent A scintillator-bar array, hereafter named as scintillator wall, is combined with the TPC for pion measurements. It is attached to the back side of the TPC, as indicated on the sketch of \figref{fig:r3b}. It provides (i) a trigger for the DAQ, and (ii) a start (time) signal for drift-time measurement in the TPC.
        The wall consists of 16 plastic scintillator bars of type EJ-200~\cite{EllenTechnologies2025} (see \figref{fig:pw}), wrapped with aluminized Mylar foil for light reflection and a black vinyl light-tight layer to prevent crosstalk between neighboring bars. To further ensure light tightness, the full assembly (w/o electronics components) was wrapped in aluminum foil and covered using a 3D printed mechanical structure. These 16 bars, each with a length of 250~mm and a width of 23~mm, fully cover the exit window of the TPC. These bars have a thickness of 4~mm, selected such that the mean energy deposition of pions originating from hypernuclei decay is 1~MeV in the scintillating material, corresponding to a light yield of $\sim$10,000 photons.\\

        \noindent To guarantee the functionality of the scintillator wall in strong magnetic fields of the GLAD dipole magnet, Silicon Photomultipliers (SiPMs) are used for the detection of the scintillation light. Two Hamamatsu SiPMs of the S13360 series~\cite{Hamamatsu2025} are connected to the top side of each scintillator bar using optical-grade silicone grease. The bars feature a trapezoidal section at this edge, serving as a light guide. Given that the detector is not required to provide the hit position information along the bar, the scintillation signal is extracted from the top side only (see \figref{fig:pw}), allowing for a compact design. A more stable connection is expected by using silicone pads at a subsequent stage. Each SiPM has dimensions of 3$\times$3~mm$^2$ and a pixel pitch of 50~\textmu m, comprising a total of 3,600 pixels. The photon detection efficiency of these SiPMs is optimized to be maximal in the wavelength range of 400-510~nm, corresponding to the emission spectrum range of the scintillator.\\

        \noindent To model the detector response, simulations including optical photon tracking have been performed based on the Geant4 toolkit~\cite{Agostinelli2003}. In particular, the GODDeSS library extension has been used~\cite{Dietz-Laursonn2017} to simulate photon emission and transport to SiPMs. The geometry and materials of the TPC and scintillator wall have been considered, incorporating the GLAD field map. Pions were generated from hypernuclei decay. Specifically for the scintillator wall, features of both the scintillation and wrapping materials, e.g., refraction, optical surface, and reflection, have been taken into account, while for SiPMs 100\% detection efficiency has been assumed. The simulated data have been adjusted to account for the photon detection efficiency of the SiPMs, which encompasses both the wavelength-dependent quantum efficiency and the geometric fill factor~\cite{Hamamatsu2025}.\\

\begin{figure}[h!]
        \centering
        \includegraphics[trim=0cm 0cm 0cm 0cm,clip,width=0.75\textwidth]{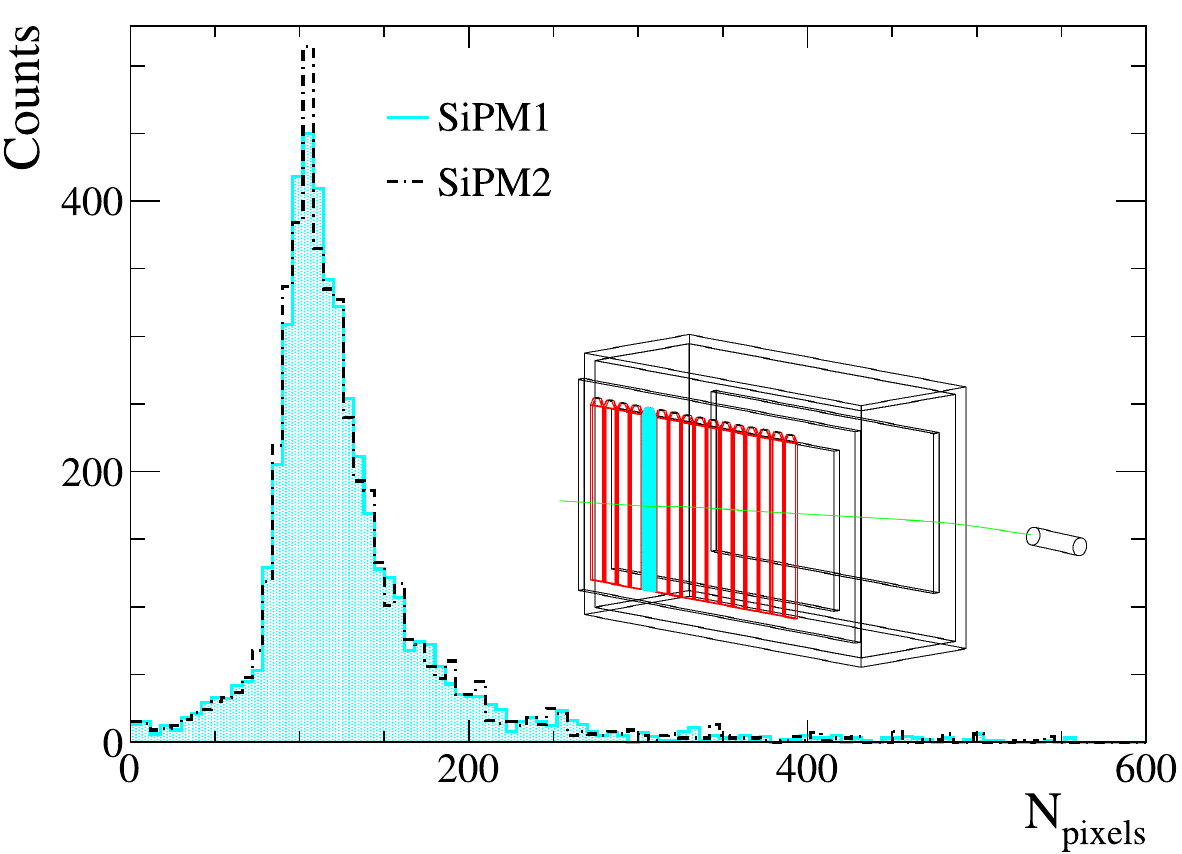}
        \caption{Simulated detected photon number ($\textrm{N}_{\textrm{pixels}}$) after the energy loss of pions coming from a hypernucleus decay in a module of the scintillator wall. The histograms correspond to the two SiPMs attached to the bar. The inset illustrates the trajectory of one pion (green) with a kinetic energy of 0.5~GeV.}
        \label{fig:pion}
        \end{figure}

        \noindent The resulting photon number ($\textrm{N}_{\textrm{pixels}}$) spectrum for pions traversing a scintillator bar with realistic kinematics is shown in \figref{fig:pion} for the two SiPMs of the relevant bar. The mean number of pixels per SiPM is above 100 photons, significantly greater than the typical dark current threshold of 10-20 photons. 

    \subsection{Electronics}
\begin{figure}[h!]
        \vspace{-10pt}
        \centering
        \includegraphics[trim=0cm 3cm 0cm 4cm,clip,width=0.75\textwidth]{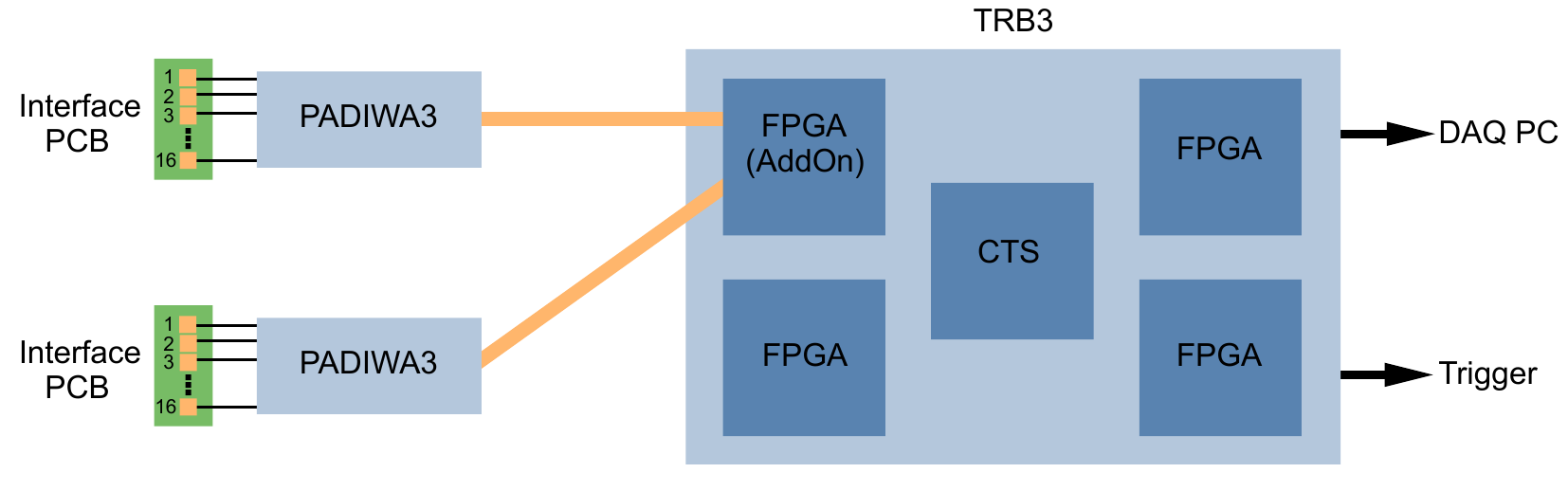}
        \caption{Schematics of the scintillator wall electronics: signals from the SiPMs, soldered on an interface PCB, are sent to the front-end PADIWA3 boards and are then transmitted to an FPGA in the TRB3 board. The CTS executes the trigger logic, and the data is sent to the DAQ PC, in addition to a trigger output that can be sent to the TPC. }
        \label{fig:trb}
        \end{figure}

        \noindent The SiPMs are soldered to two interface PCBs, each of them containing 16 SiPMs positioned with the pitch corresponding to the scintillator bar footprint. On the other side of the PCBs, MML cables direct the analog signal of the individual SiPMs to the front-end electronics. A schematics of the read out chain is shown in \figref{fig:trb}. Each PCB powers all 16 SiPMs, which operate at a recommended voltage of $56\pm5$~V, through a single LEMO cable.\\

        \noindent For high time resolution, signals are read with the TRB3-based (TDC Readout Board)~\cite{Traxler2011} system, used as a TDC (Time-to-Digital Converter) together with a front-end electronic module, the PADIWA3~\cite{Cardinali2014} board. The PADIWA3 board amplifies and discriminates the signals, i.e., converting the analog signals to digital signals. The signal width is encoded into the timing of differential signals and subsequently sent to the back-end TRB3 TDC via long ribbon cables. Two PADIWA3 boards are used for the scintillator wall, each powered individually with 5~V. To minimize the influence of noise on the analog signals, the PADIWA3 boards are placed in proximity to the SiPMs. The TRB3 uses Low-Voltage Differential Signal (LVDS) input buffers of an FPGA to realize leading edge discrimination for 16 input channels. The time information is extracted from the leading edge LVDS signal, and the amplitude from the time difference between the leading and falling edges, i.e., the time-over-threshold.\\

        \noindent The TRB3~\cite{Traxler2011} is a high-precision TDC platform for time measurements implemented in FPGAs, developed at GSI. The TRB3 board is equipped with 5 FPGAs and requires 15~V power supply. Four peripheral FPGAs can be programmed to provide 64 TDC channels (plus one reference channel), and the fifth FPGA serves as a CTS (Central Trigger System). The central FPGA coordinates the peripheral ones and communicates with the DAQ system DABC~\cite{Adamczewski-Musch2012}. In practice, up to 3 PADIWA3 front-end boards can be connected to each of the peripheral FPGAs, using a small AddOn-PCB plugged onto the TRB3, resulting in 48 channels. After digitization, the data is collected by the CTS and sent via a gigabit Ethernet cable to the DAQ PC. The internal time precision of the system was demonstrated to be very high, reaching values as low as 8~ps (RMS)~\cite{Ugur2016} for leading-edge measurements of a single channel. \\

        \noindent The scintillator wall provides a trigger signal to the HYDRA TPC. The trigger logic is executed in the CTS of the TRB3 board and can send a trigger output in the form of a TTL signal. In the current trigger scheme, a trigger signal is sent when two SiPMs of the same module register a hit, i.e., a coincidence between two channels connected to one bar and a logical OR between all pairs of channels. The coincidence between two channels reduces the number of non-physical hits and improves the trigger selectivity. Additionally, since the scintillator bars are read out only on one side, high thresholds are applied on the PADIWA3 channels. The TDC channels can process bursts with a maximal rate of 50~MHz, and store measurements of up to 63 hits before a readout. The maximal readout trigger rate is about 700~kHz, depending on the configuration used and network size.

    \subsection{Validation measurements}

    \subsubsection{Time precision}

        \noindent As the scintillator wall provides the start signal for drift-time measurement in the TPC, it is of importance to validate the high time precision of the TRB3-based system. A TDC is implemented on each of the peripheral FPGAs to determine the hit times. The TDCs contain coarse and fine time counters. The coarse time counter is incremented with a 200~MHz clock, i.e., all steps of the fine counter sum up to 5~ns. As the fine counter steps are not fixed, they have to be determined to guarantee precise time measurements. For this purpose, as a first step, a time calibration of the TDCs is performed. To measure the internal time precision of the TDC we used a pulser signal, which was split and sent to two PADIWA3 channels. By measuring the time difference between the signals arriving at the two channels, a time resolution of $\sim15$~ps ($\sigma$) was measured for a single channel, consistent with the designed value. Following that, the time resolution in a scintillator bar is expected to be better than 100~ps.  

    \subsubsection{Cosmic rays}

\begin{figure}[h!]
        \centering
        \includegraphics[trim=0cm 0cm 0cm 0cm,clip,width=0.75\textwidth]{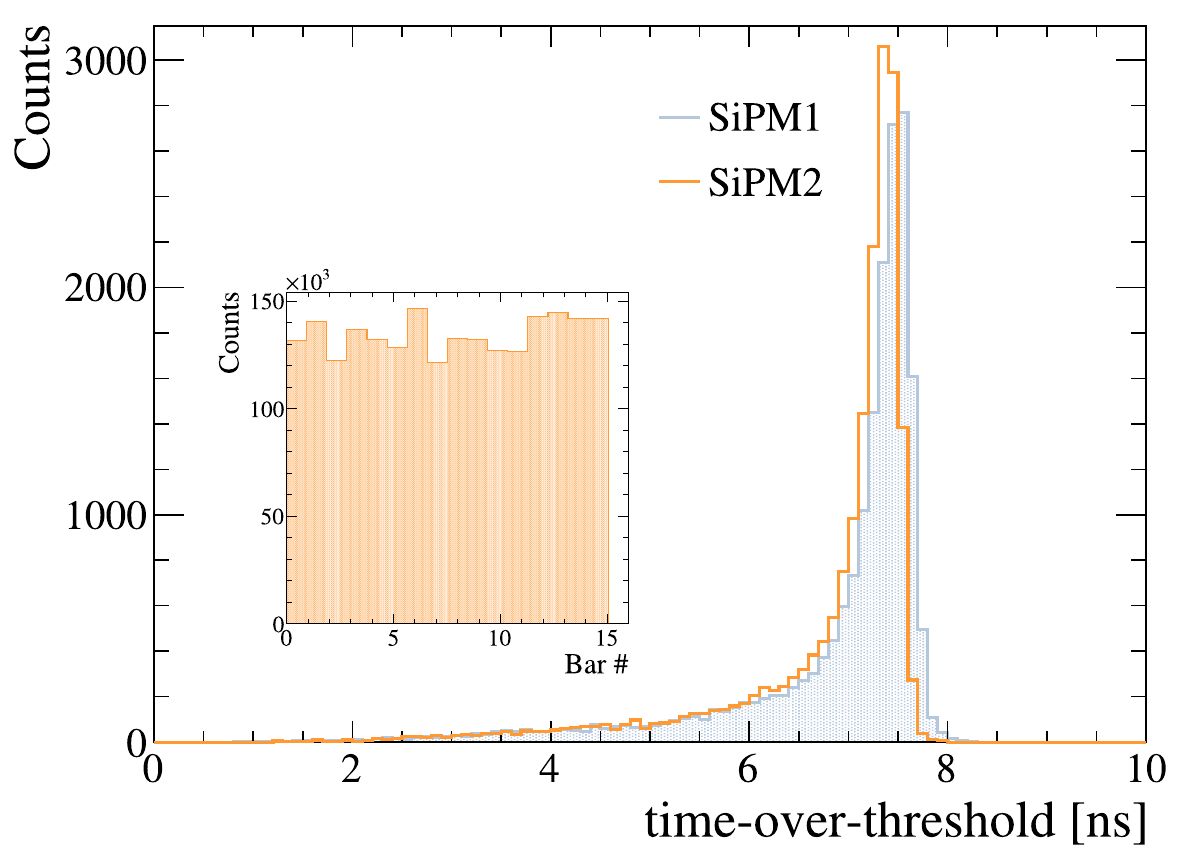}
        \caption{Data from cosmic muon measurement: time-over-threshold distribution of two SiPMs from one bar in the scintillator wall. The inset shows the count rate for all bars, which is approximately constant.}
        \label{fig:cosmics}
        \end{figure}

        \noindent To validate the operation of the scintillator wall, cosmic muons have been measured. These are minimum ionizing particles with a constant energy deposition in the scintillation material of 2~MeV$\cdot{\rm cm}^{-1}$. As the thickness of the bars is 0.4~cm, this corresponds to energy deposition of roughly 1~MeV, similar to the signal expected from pions. \Figref{fig:cosmics} shows the time-over-threshold distribution measured for two channels in one bar of the scintillator wall. A clear peak is seen well above the threshold, attributed to cosmic muons. The small differences in the shape of the distribution between the two channels may arise from the uniform voltage (54.7~V) applied to all SiPMs, which can have a slight gain difference. The tail towards lower values originates from the hit position along the bar (which can not be determined as the bars are read out only on one side) that leads to different measured times and low-energetic terrestrial background. Distributions for other channels show similar characteristics with an almost constant count rate (see inset in \figref{fig:cosmics}), since no dependence on the bar position is expected for cosmic muons.

    \subsubsection{Rate dependence}

        \noindent To test the rate capability and the dependence on the intensity of the incoming beam, an in-beam measurement was performed at the R$^3$B setup. The scintillator wall was attached to the HYDRA TPC that was placed inside the GLAD dipole magnet, operated at 1.6~T. A $^{12}$C beam at an energy of 1.9~GeV/nucleon was sent onto a 10~cm thick cylindrical graphite target to induce ion-ion collisions. Since the GET-based electronics was used for the TPC (see \secref{sub:get}), it was not feasible to reconstruct tracks in the TPC due to the high trigger rate and multiplexing scheme. Therefore, the main objective was to test the operation of the scintillator wall and, in particular, to validate the trigger rate, which was based on simulations solely. This is important when considering ion back-flow in the TPC, where the rate of charged particles traversing the drift region leads to space charge density within the active volume. With simulated rates and the hybrid amplification in the TPC, the IBF rate is estimated below 1\% (see \secref{sub:amp}), minimizing electric field distortions.\\

\begin{figure}[h!]
        \centering
        \includegraphics[trim=0cm 0cm 0cm 0cm,clip,width=0.75\textwidth]{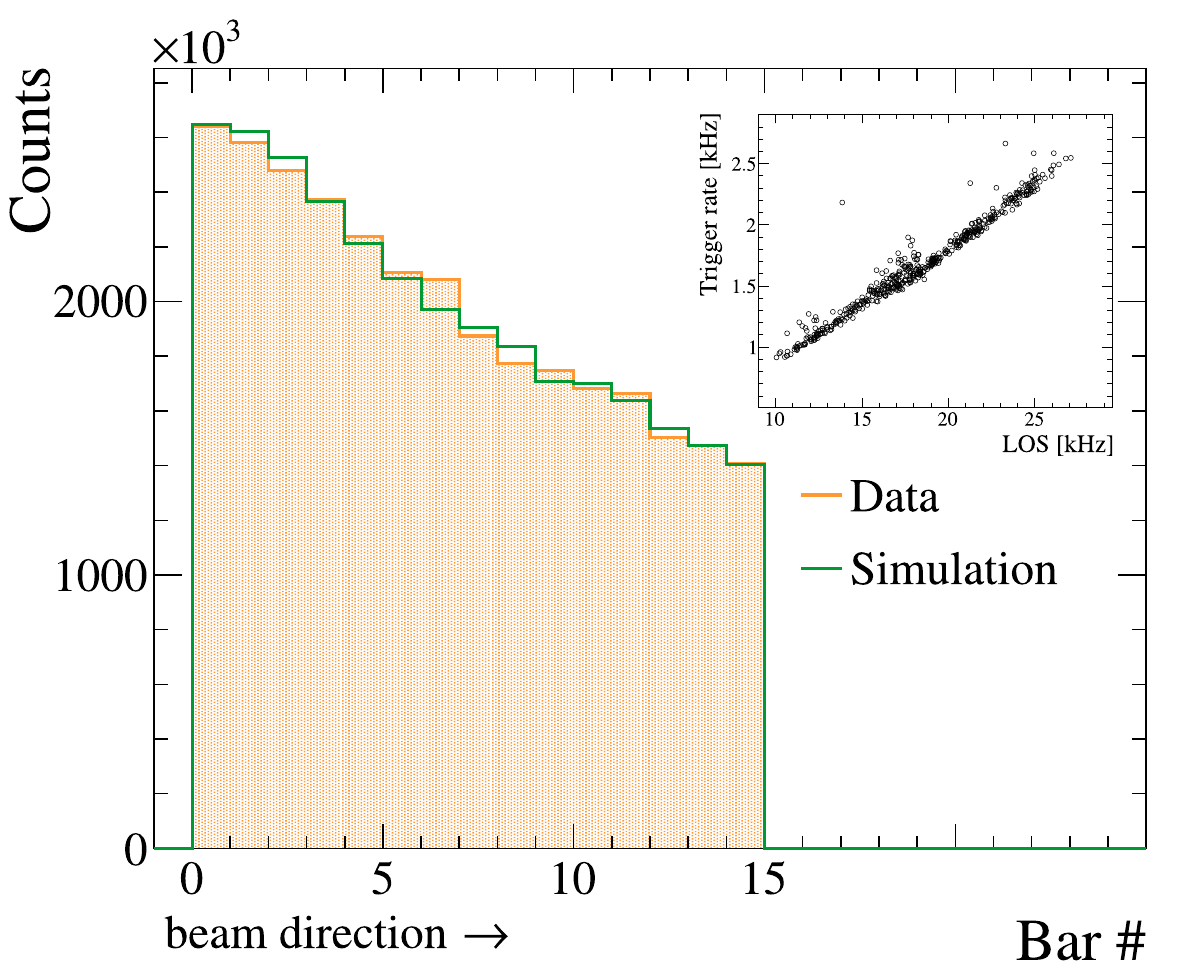}
        \caption{Comparison of measured and simulated bar distribution in the scintillator wall from $^{12}$C+$^{12}$C collisions. Bar \#15 was excluded from the analysis due to a problem in the optical coupling. The inset shows the measured trigger rate vs. the incoming beam rate measured by the start detector of the R$^3$B setup LOS. }
        \label{fig:beam}
        \end{figure}

        \noindent During the beam time, the operation of the scintillator wall was stable for all beam intensities. The incoming beam rate was measured by the start detector of the R$^3$B setup LOS and was correlated with the measured trigger rate in the scintillator wall (see inset in \figref{fig:beam}). As an example, at the incoming rate of 10~kHz, the measured trigger rate was $\sim1$~kHz. For comparison, $^{12}$C+$^{12}$C collisions were simulated in Geant4 using the INCL++ model~\cite{Leray2013}, resulting in an expected rate of 1.17~kHz, consistent with the measured one. Furthermore, when comparing measured distributions, a good agreement was found as shown in \figref{fig:beam} for the bar distribution. 

\section{Characterization Measurements}
    \subsection{Gain characterization}
        \label{sec:gain}
        \noindent The gain of the amplification stage is determined by the HV settings of the GEM plus Micromegas hybrid detector, as well as the drift and transfer fields. We define the total effective gain as the ratio between the current on the DLC layer ($I_{\mathrm {amplified}}$) to the primary current induced by ionization ($I_{\mathrm{primary}}$) 
        \begin{align}
        \mathrm{Gain}\:=\:\frac{I_{\mathrm {amplified}}}{I_{\mathrm {primary}}}.
        \end{align}

        \noindent In the following, gain curves are extracted for the \ArCOtwo gas mixture maintained slightly above atmospheric pressure using a $^{55}$Fe source, which emits mainly 5.9~keV X-rays at a rate of approximately 100 kHz in the TPC. The source was placed in front of the entrance window of the TPC at a distance of about 20~cm. \\

        \noindent For these measurements, the DLC layer was connected to a floating multichannel picoamperemeter~\cite{Utrobicic2015}. The current with ($I^{\rm w/}_{\rm DLC}$) and without ($I^{\rm w/o}_{\rm DLC}$) the source was measured such that $I_{\rm amplified}  =  I^{\rm w/}_{\rm DLC} - I^{\rm w/o}_{\rm DLC}$. 
        To extract $I_{\rm primary }$, the rate of primary electrons on the micromesh was determined by integrating the energy spectrum.
        \begin{table}[h!]
        \caption{High voltage settings for gain characterization measurements shown in \figref{fig:gain} panels (a)-(d). The brackets represent ranges of potentials and fields.}
        \centering
        \setlength{\tabcolsep}{2pt}
        \begin{tabular}{ l c c c c }
            \hline
            Region & (a) & (b) & (c) & (d) \\ 
            \hline
            $\mathrm{\Delta U_{DLC}}$ [V] & [380,440] & 440 & 440 & 440 \\
            ${E_{\mathrm {transfer}}}$ [V/cm] & 1000  & [200,2400] & [0,2000] & 1000 \\
            $\mathrm{\Delta U_{GEM}}$ [V] & 350 & [310,350]  &[230,430]   & 300, 350 \\    
            ${E_\mathrm{drift}}$ [V/cm] & 220 & 220 & 220 & [40,220] \\
            \hline
        \end{tabular}
        \label{tab:HV-setting}
        \end{table}
 
        \noindent The TPC gain characterization is performed by scanning the potential applied in the DLC layer ($\rm{U_{DLC}}$), the transfer field ($E_{\rm{transfer}}$), the GEM voltage ($\rm{\Delta U_{GEM}}$ $ = \rm{U_{bottom}-U_{top}}$), and the drift field ($E_{\rm{drift}}$). Four sets of measurements were taken with different HV settings, as summarized in Table~\ref{tab:HV-setting}.\\
        \\
        \\
        \begin{figure*}[h!]   
        \centering
        \includegraphics[width=\textwidth]{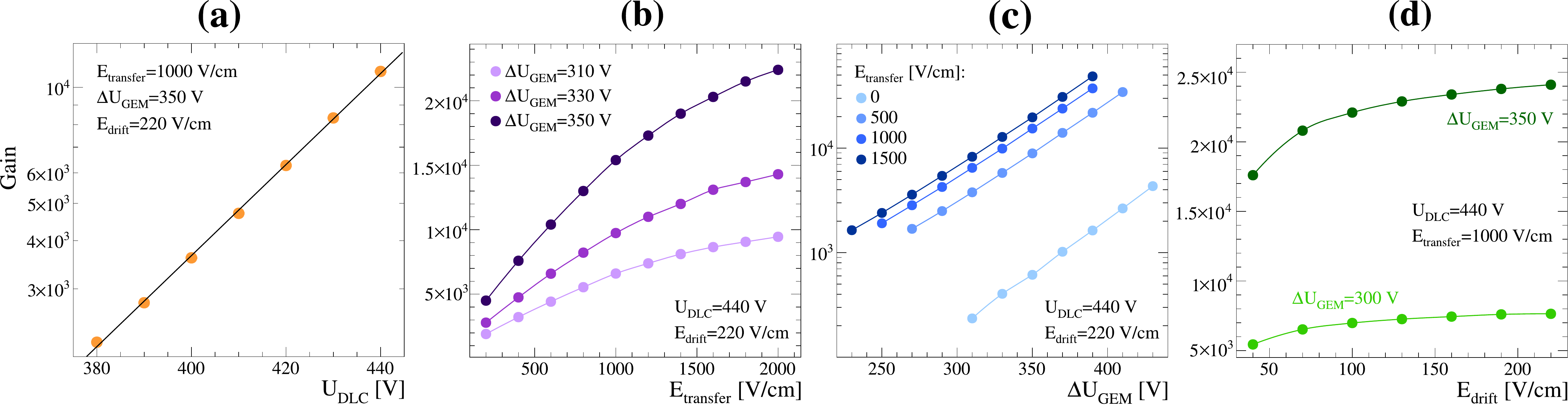}
        \caption{Measured effective gain curves as a function of different parameters. (a) Micromegas - as a function of the potential on the DLC layer ($\rm{U_{DLC}}$). Data is shown in logarithmic scale, together with a linear fit (black). (b) Transfer region - as a function of the field strength ($E_{\rm transfer}$). The curves correspond to different GEM potentials. (c) GEM - as a function of the potential difference between bottom and top sides ($\rm{\Delta U_{GEM}}$). Data is shown on a logarithmic scale. The curves represent different transfer field strengths. The curve for $E_{\rm transfer}=2000$~V/cm is not shown as it is almost identical to that at 1500~V/cm. (d) Drift region - as a function of the drift field ($E_{\rm drift}$) for two GEM potentials. In all panels, the error bars associated with the data points are smaller than the symbol size and are, therefore, not shown. }
        \label{fig:gain}
        \end{figure*}
        
        \noindent The resulting gain curves shown in the four panels of \figref{fig:gain} follow expectations from the well-established charge amplification and transfer processes in GEMs (see \cite{Bachmann1999}, for example). In the Micromegas (panel a) and GEM (panel c) voltage scans, the number of avalanche electrons generated per primary electron increases exponentially with the amplification field. The transfer field values are too low to allow for any amplification in the transfer gap. However, an increasing transfer field $E_{\rm transfer}$ allows for a more efficient extraction of electrons from the GEM towards the Micromegas stage. In this case, the gain increases approximately linearly with the field strength up to $\sim$1000~V/cm, where it starts to saturate, as shown in panel (b). This trend is independent of the voltage applied to the GEM, implying that the electron extraction is not influenced by the amount of electrons produced in the GEM region. It is further illustrated in (c) for the GEM. Finally, the drift field, in the measured range, does not influence the amplification process, while it impacts the electron collection efficiency, thus the gain. In weak electric fields, the probability of electron attachment increases significantly, while at extremely low strengths, electron-ion recombination occurs. Additionally, in weak fields, the increased diffusion of electron clusters can lead to their absorption by the GEM. These processes lead to electron losses. Panel (d) shows the gain curve for the drift region. By increasing the field strength the fraction of electron loss is reduced, and hence, the gain increases until the curve flattens, indicating that all drifting electrons are collected within the GEM holes. \\

        \noindent A gain of $\mathcal{O}(10^4)$, as used in standard operations with HYDRA and in the following laser measurements (see \secref{sec:measurement-with-laser}), can be achieved with the following set of settings: $\rm{U_{DLC}}=400$~V, $E_{\rm transfer}=1000$~V/cm, $\rm {\Delta U_{GEM}=380}$~V, and $E_{\rm drift}=220$~V/cm. 
    
    \subsection{Commissioning with Laser system} \label{sec:measurement-with-laser}
    \subsubsection{Readout system}\label{sub:get}

\begin{figure}[h!]
        \centering
        \includegraphics[trim={0cm 0cm 0cm 0cm},clip,width=0.75\textwidth]{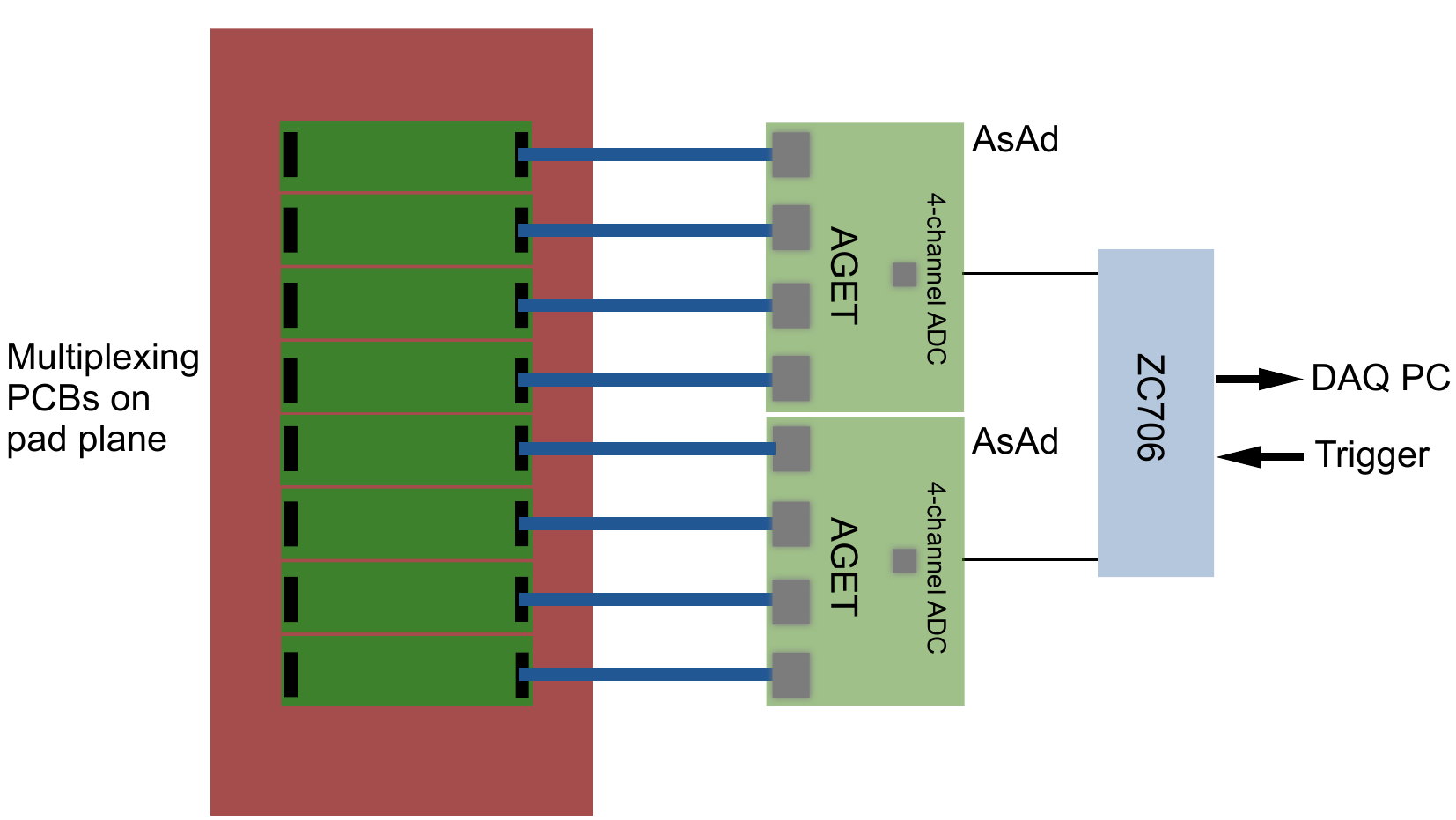}
        \caption{Schematics of the TPC electronics used for the commissioning: signals from the pad plane are multiplexed by 8 PCBs, sent to the AGET chips on the front-end AsAd boards and are then transmitted to the concentrator ZC706 board. For illustration, only the right side of the TPC is shown. The left side is read out in the same way. } 
        \label{fig:get}
        \end{figure}

        \noindent A 1024-channel GET-based electronics readout was used for measurements with the laser system. GET (General Electronics for TPCs)~\cite{Pollacco2018} is a generic electronics system for TPCs and nuclear physics instrumentation. A schematic of the readout chain is shown in \figref{fig:get}. \\

        \noindent The front-end electronics is based on the AGET (ASIC for GET) chips~\cite{Pollacco2018}. Each AGET chip has 64 input signal channels and 4 additional fixed-pattern noise (FPN) channels. The FPN channels are not connected to the detector but treated in the same way as the signal channels, and can be used to determine the intrinsic noise level and baseline. For each channel, the main components are the charge sensitive pre-amplifier (CSA), the shaper, the analog memory, and the discriminator. The CSA has four different gain settings for dynamic ranges of 120~fC, 240~fC, 1~pC, or 10~pC, covered by 4096 ADC channels. The shaper stage is a filter with a peaking time that can be selected from 16 values in the range between 70~ns to 1~$\upmu$s. The output signal of the shaper is sampled in time and stored into the analog memory, to be further digitized by an ADC at a 25 MHz frequency. The memory contains 512 memory cells that can store the signal with a write frequency ranging from 12.5 to 100~MHz. For the measurements presented in this chapter a gain of 120~fC, a peaking time of 232~ns, and a write frequency of 12.5~MHz were used. \\ 

        \noindent The AsAd (ASIC and ADC) board includes four AGET chips and a four-channel ADC. It can process up to 256 analog input signal channels plus 16 FPN channels, digitize the samples stored in the analog memory, and send these to the concentration board. The output contains the waveform samples for all channels, the ADC values for the 512 time bins (cells). \\

        \noindent The concentration of events is done at the back-end system developed by the University of Warsaw. It is based on a commercial Xilinx Zynq-7000 module, implemented on ZC706~\cite{ZC706evaluationboard2025} evaluation board with the z-CoBo firmware~\cite{Mwiokandothers2023}, that can support two AsAd boards. For the test measurements, 4 AsAd boards were used, which were stacked together and connected to two ZC706 boards that were triggered by the laser system. The output from these boards is sent to a PC where the GET DAQ is running on. The trigger rate of the system is limited to a maximum of 1~kHz. As presented in \secref{sec:req}, high beam intensities are required in the future R$^3$B experiments with HYDRA, and therefore, a high-rate capability readout based on the VMM3a ASIC~\cite{Iakovidis2020} is developed.

    \subsubsection{Multiplexing}
\begin{figure}[h!]
        \centering
        \includegraphics[width=0.75\textwidth]{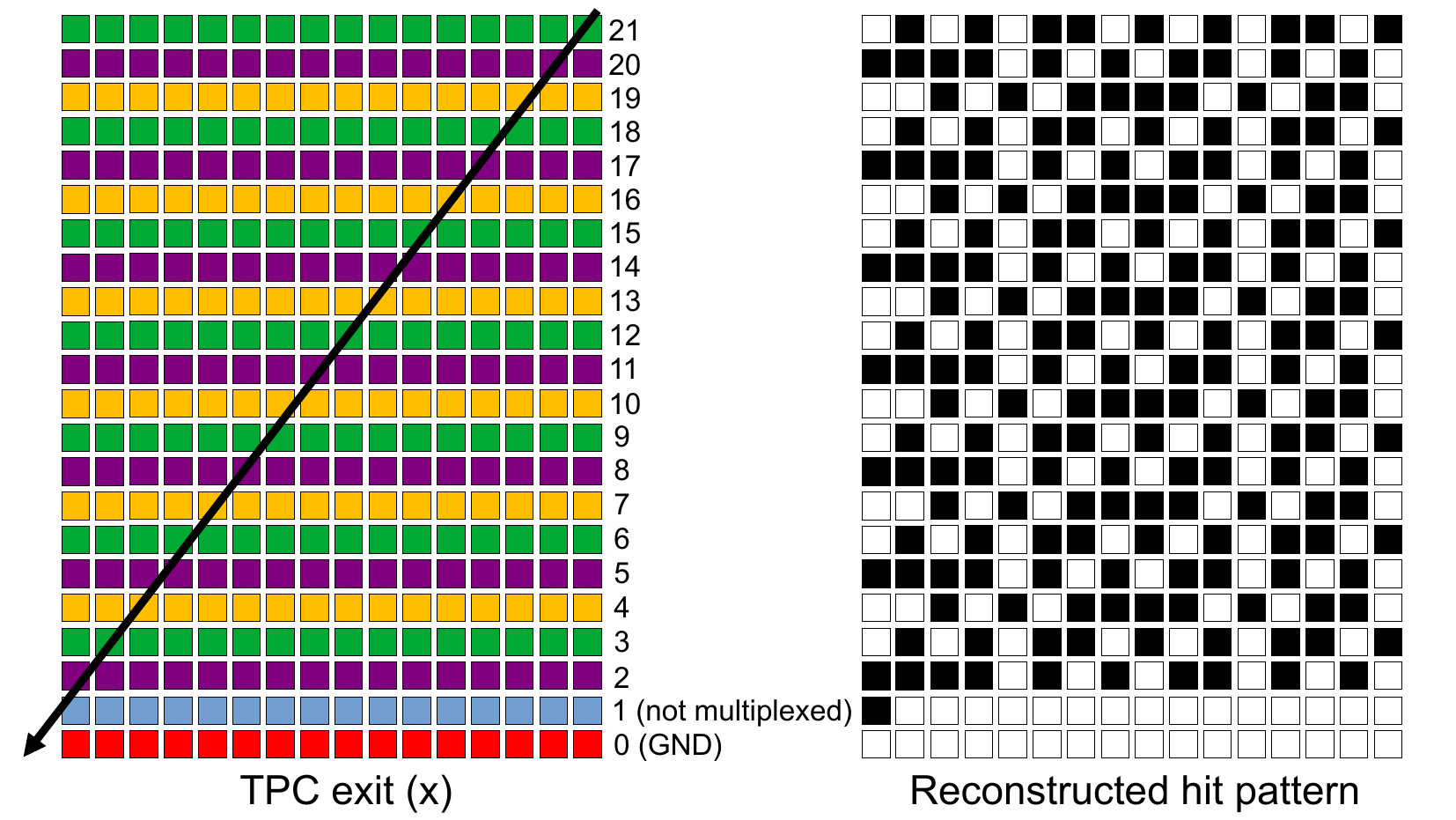}
        \caption{Multiplexing scheme in one half (along X direction) of the PCB (left). The first row is grounded, while pads in the second row are not multiplexed and can be used as a reference point for track reconstruction. In the remaining rows, 7 (purple/green) or 6 (yellow) pads are grouped. For illustration, an arbitrary laser track is shown together with the resulted reconstructed hit pattern (right). The pad in the second row is used to put a constraint on the track reconstruction in the analysis.}
        \label{fig:multiplexing}
        \end{figure}

        \noindent To read out the total number of 5632 pads in the TPC with 1024 electronic channels, several pads are grouped and connected to one electronics channel. This is implemented by the design of multiplexing PCBs which are connected on top of the pad plane. Each multiplexing board covers a pad region of 44$\times$16 in the XZ plane, such that 8 PCBs are used to cover the full TPC. Along the 44 rows, the first and last ones are grounded. The second and second-to-last rows of pads are not multiplexed and are used as reference points in the analysis. In the remaining rows pads are multiplexed into groups of 6 or 7, as illustrated in \figref{fig:multiplexing}. Strictly speaking, the grouping pattern is not a 2D generic multiplexing (as done in e.g.~\cite{Bouteille2016}), where combinatorics allow to recover exactly the lost information. Yet, here we refer to it as multiplexing, since, for single laser tracks, the full information can be recovered, as explained below. 128 output signals from each multiplexing board are transferred to two AGET chips via 1~m long SAMTEC ERCD ribbon cables, with a capacitance of 88.58 pF and a characteristic impedance of 50 ${\rm \Omega}$, connected on both sides of the PCB.
   
    \subsubsection{Results}
        \noindent Measurements were performed with the UV laser source and mirror assembly installed inside the TPC (see \secref{sec:laser}). The TPC was operated with the \ArCOtwo gas mixture with a gain of $\mathcal{O}(10^4)$ (see Table~\ref{tab:voltages} in \secref{TPC:HV}). The laser beam entering the TPC is split and reflected by 3 micro-mirror bundles placed at different angles and heights (see \figref{fig:laser}), generating laser tracks in the active volume. These tracks ionize the gas via two-step ionization, generating signals on the pad plane. \\ 
    
        \noindent The laser source sends a trigger signal at a frequency of 20~Hz to the back-end ZC706 board. Once a trigger is received, the electronic system records the waveforms for all channels within a pre-defined time window. In the first step, the electronics parameters were optimized for the laser measurements. For high sensitivity, the lowest range parameter of 120~fC was chosen, such that each ADC channel corresponds to a charge of $\sim$180 electrons. For the \ArCOtwo gas mixture, the calculated minimum drift time, for electrons to traverse the drift region of 300~mm is $\sim$20~$\upmu$s~according to Table \ref{tab:gas}. With a readout frequency of 12.5 MHz, the size of the time bin is 80~ns, and the corresponding time window is 40.96~$\upmu$s, such that the drift region is fully covered. In addition, a trigger delay of 32~$\upmu$s was implemented, which defines the latency time for the system to start the signal processing after a trigger was received. \\
    
        \begin{figure}[h!]
        \centering
        \includegraphics[width=0.75\textwidth]{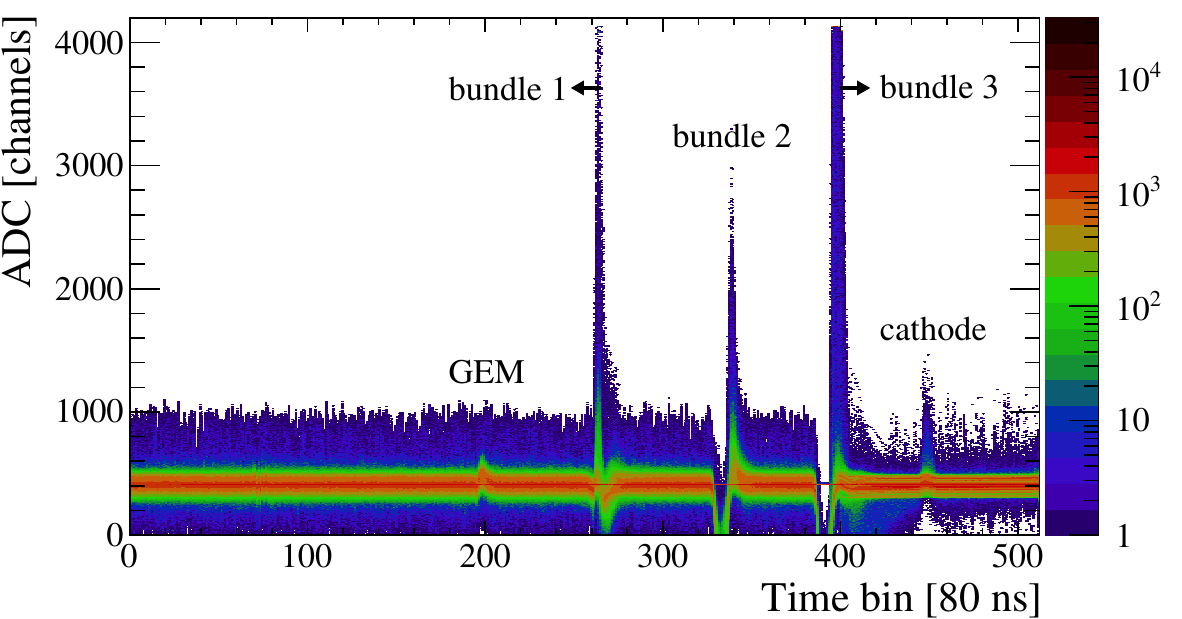}
        \caption{Raw waveforms for 2069 events in one AGET chip. The peaks observed in the data (in increasing time order) are attributed to laser reflections on the GEM foil, laser tracks reflected by the three micro-mirror bundles, and reflections from the cathode.}
        \label{fig:peaks}
        \end{figure}
    
        \noindent As the micro-mirror bundles are placed at different heights along the drift direction, the produced electrons of corresponding laser tracks exhibit different drift times. Therefore, three distinct signals are expected from the laser tracks that are reflected by the micro-mirror bundles. \Figref{fig:peaks} shows an example of the raw data measured in one AGET chip, where three main peaks attributed to the laser tracks are observed. Two additional lower peaks can be identified at shorter and longer drift times. The former is associated with signals originating from the GEM (shortest drift distance), where laser reflection onto the bottom layer of the GEM can lead to the photoelectric effect in the material. Similarly, the peak corresponding to the longer drift time can be associated with signals originating from the cathode (largest drift distance). As the cathode-to-GEM distance corresponds to the full drift region of 300~mm, the data is used to extract the drift velocity. The peak amplitudes and times are determined by identifying the time-bin with the maximum ADC value in a specific region, resulting in a drift velocity of 1.50 $\pm$ 0.01 cm/$\upmu$s. For comparison, the calculated value (see Table~\ref{tab:gas}) is 1.56~cm/$\upmu$s, with a previously measured one under $E_d$ = 220 V/cm of 1.57~ cm/$\upmu$s \cite{Zhao1994}.\\
 
        \noindent After noise subtraction, the peak amplitude and time for each candidate laser track are extracted, and the corresponding hit patterns on the pad plane are reconstructed by mapping the channels according to the multiplexing scheme. An example for a candidate track from the first micro-mirror bundle is shown in panel (a) of \figref{fig:hits}. To overcome the difficulties imposed by the multiplexing scheme for the track reconstruction, a de-multiplexing method has been implemented. To do so, the information from the non-multiplexed channels connected to the outermost row of pads on each side of the pad plane (along the x-axis) is used. These pads provide the exit position of the laser track and are used as constraints in the reconstruction algorithm, performed iteratively in the following steps:
        \begin{enumerate}
        \item Find the hit cluster corresponding to the laser track among all non-multiplexed pads, i.e., with index X=1 in \figref{fig:hits}. \\
        
        \item Calculate the centroid of charge $\Bar{\rm {X}}_Z$ for each pad index Z among the cluster of non-multiplexed pads and the neighboring multiplexed pads with index ${\rm X}>1$, following
        \begin{align}
            \Bar{\rm X}_Z\:=\:\frac{\sum_{\{ij\} \in {\rm cluster}} Q_{ij}\:\cdot\:{\rm X}_{ij}}{\sum_{\{ij\} \in {\rm cluster}} Q_{ij}}\cdot \delta_{iZ},
            \label{eqn:centroid}
        \end{align}
        where $i$ and $j$ denote the Z and X indices of the pads, respectively; X$_{ij}$ and $Q_{ij}$ denote the X coordinate and the ADC value of the pad ($i$,$j$), respectively. \\
        
        \item Perform a linear fit of the centroid points $\{Z,\Bar{\rm X}_Z\}$. In the single track of \figref{fig:hits}, the considered pads for this initial cluster related to the exit location of the laser track are highlighted by a black box. The linear fit is therefore performed over 6 centroid points for $Z \in [79,84]$.\\
             
        \begin{figure}[h!]
        \centering
        \includegraphics[width=0.75\textwidth]{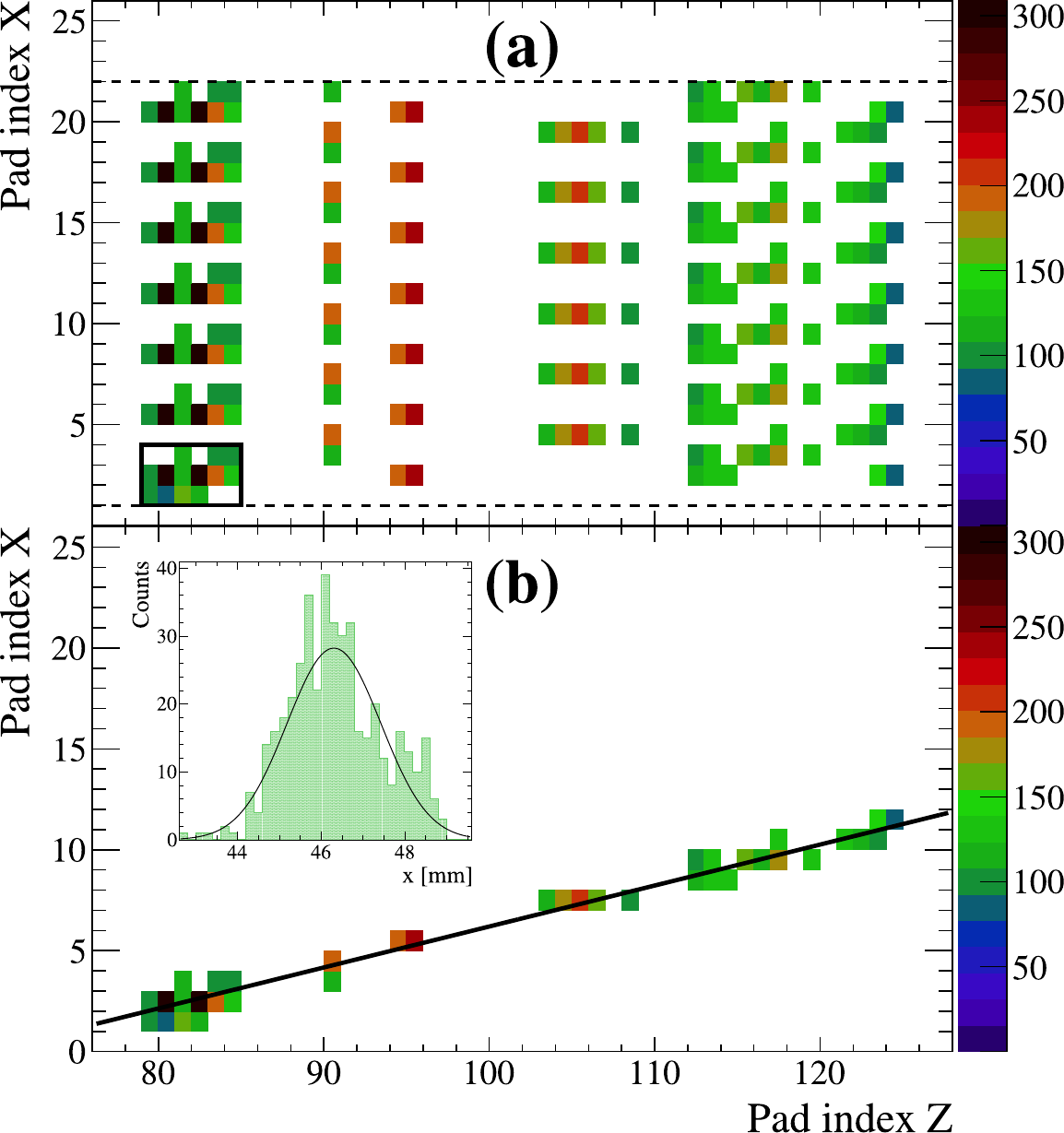}
        \caption{(a) Initial reconstructed hit pattern of one candidate laser track in the XZ pad plane following the multiplexing mapping. After hits in the non-multiplexed pads are identified (Pad index X=1), these pads and their neighboring multiplexed pads (black rectangle) are selected, and the de-multiplexing method is applied, resulting in a straight reconstructed track (b). The black line represents a linear fit. The color code indicates the amplitude of the hits above the baseline in ADC channels. The inset shows the extrapolated X coordinate of the micro-mirror from $\sim$400 reconstructed tracks, together with a Gaussian fit (black).  }
        \label{fig:hits}
        \end{figure}
        
        \item Extrapolate the linear function to the index Z of the next cluster of multiplexed pads with non-zero hits (Z=91 in \figref{fig:hits}). Select the pad cluster with the minimal distance to the line, and calculate the centroid of charge $\Bar{\rm X}_Z$. \\
        
        \item Perform another linear fit including both the new centroid of charge and those previously identified. \\
        
        \item Repeat steps 4-5 until the linear function reaches the last Z index with hits as illustrated in panel (b) of \figref{fig:hits}.
        \end{enumerate}
    
        \noindent With this method, laser tracks were reconstructed successfully in the XZ plane. In addition, by considering the drift time information, the vertical distance along the y-axis is extracted. The spatial resolution for track reconstruction can be extracted by calculating the dispersion of the centroid of charge from the reconstructed track. The resulted resolution ($\sigma$) for laser tracks originating from the first micro-mirror bundle in the XZ pad plane is 0.46~mm and 0.61~mm in the vertical drift plane. 
        It should be noted that, in this first characterization, the variations in the gain of the electronics and the amplification structure were not corrected for. \\
    
        \noindent To assess the performance of the method, the fit functions are extrapolated to the pad index Z corresponding to the micro-mirror’s position to determine its X coordinate. The inset in \figref{fig:hits} shows the resulted distribution, with a mean value of 46.3$\pm$0.1~mm and standard deviation of 1.1$\pm$0.1~mm. Given the uncertainty for a specific micro-mirror within the bundle responsible for reflecting the laser track, and considering that the bundle center is aligned with the pad plane center, the theoretical coordinate is 44$\pm$1.5~mm, consistent with the measured one within two standard deviations.

\section{Conclusion}
    \noindent We report on the HYpernuclei-Decay at R$^3$B Apparatus (HYDRA) pion tracker composed of a time-projection chamber (TPC) and a scintillator wall for timing and trigger purposes. HYDRA is intended to be used inside the large-acceptance GLAD dipole magnet of R$^3$B at GSI-FAIR to measure pions from the weak decay of light hypernuclei produced from ion-ion collisions at energies in the laboratory frame above the production threshold of 1.6 GeV/nucleon. HYDRA is designed to achieve a 1\% momentum resolution in the case of a homogeneous magnetic field of 2 T for the expected kinematical conditions, corresponding to an invariant-mass resolution of 1.5 MeV/c$^2$ ($\sigma$). \\
  
    \noindent The TPC offers an active volume of 6.8 dm$^3$ above a detection plane composed of 128$\times$44 pads, each with dimensions of 1.9$\times$1.9 mm$^2$. A two-layer wire field cage encloses the drift region of the TPC together with a copper cathode and a GEM foil. A hybrid amplification stage comprising the GEM and a resistive Micromegas is employed, aiming at an ion back-flow of $\sim$1\%.  A laser system, incorporating micro-mirror bundles, is embedded within the TPC to generate laser tracks inside the active region for drift velocity calibration and control of the tracking performances.  The TPC is designed to work at pressures slightly above atmospheric pressure and is enclosed by thin aluminized Kapton entrance and exit windows to minimize multiple scattering of incoming pions. The gain performance of the TPC was characterized using a $^{55}$Fe X-ray source. Subsequently, the first laser measurements with the TPC were successfully performed with a GET-based readout and a 6/7-fold multiplexing of the pads.\\
    
    \noindent  Combined with the TPC is a scintillator-bar array that provides a trigger as well as a start time signal for drift-time measurement. It consists of 16 plastic scintillator bars equipped with silicon photomultipliers for light detection and read out using TRB3-based electronics. The high time precision of the system was validated, and the operation of the scintillator wall was tested successfully via cosmic rays and in-beam measurements.
    
\section*{Acknowledgements}
    \noindent This work was supported by the Helmholtz Forschungsakademie Hessen f{\"u}r FAIR, Germany, the German Federal Ministry of Education and Research - BMBF project numbers 05P21RDFNB and 05P24RD1, the State of Hesse within the Research Cluster ELEMENTS (Project ID 500/10.006), as well as the Alexander von Humboldt foundation. We thank warmly Christian Schmidt, head of the GSI Detector Lab, for making the clean room and laboratory space accessible for the TPC assembly and commissioning activities, Daniel Körper for his assistance during the installation of the HYDRA setup at R$^3$B, Thomas Hackler for his support with the operation of the GLAD magnet, and the GSI accelerator team for the smooth operation during the in-beam test.

\bibliographystyle{unsrt}
\bibliography{bib}

\end{document}